\documentclass{tac}

\title{The Information Flow Framework:\\A Descriptive Category Metatheory}
\author{Robert E. Kent}
\address{Ontologos, 550 SW Staley Dr., Pullman, WA, USA 99163}
\eaddress{rekent@ontologos.org}
\copyrightyear{2004}
\keywords{descriptive category metatheory, ontologies, metalanguages, institutions}

\newcommand{\rmb}[1]{{\rmfamily\bfseries{#1}}}
\newcommand{\ttb}[1]{{\ttfamily\bfseries{#1}}}

\newcommand{\tts}[1]{{\ttfamily\small{#1}}}
\newcommand{\ttf}[1]{{\ttfamily\footnotesize{#1}}}

\newcommand{\ttbs}[1]{{\ttfamily\bfseries\small{#1}}}
\newcommand{\ttbf}[1]{{\ttfamily\bfseries\footnotesize{#1}}}
\newcommand{\sss}[1]{{\sffamily\small{#1}}}

\newcommand{\ssbf}[1]{{\sffamily\bfseries\footnotesize{#1}}}

\begin{document}
\maketitle

\begin{abstract}\label{abstract}

The Information Flow Framework (IFF)\footnote{Several years ago there was a discussion thread on the SUO email list, entitled ``composing ontologies using morphisms and colimits'' and involving Michael Uschold, John Sowa, and others.
This discussion was concerned with the question of the use of category theory (versus set theory) in the IFF project as a foundation for knowledge representation and logic.
Michael Healy forwarded this question to the category email list.
This paper can be regarded as an extended answer to this question.}
is a descriptive category metatheory currently under development that provides an important practical application of category theory.
It is an experiment in foundations,
which follows a bottom-up approach to logical description. 
The IFF forms the structural aspect of the IEEE P1600.1 Standard Upper Ontology (SUO) project. 
The categorical approach of the IFF provides a principled framework for the modular design of object-level ontologies. 
The IFF represents meta\-logic, and as such operates at the structural level of ontologies. 
In the IFF, there is a precise boundary between the metalevel and the object level. 
The modular architecture of the IFF consists of metalevels, namespaces and meta-ontologies. 
There are four metalevels,
$\mathsf{Lower}$, $\mathsf{Upper}$, $\mathsf{Top}$ and $\mathsf{Ur}$,
corresponding to the set-theoretic distinctions between
the ``small'', the ``large'', the ``very large'' and the ``generic'', respectively.
Each metalevel services the levels below by providing a metalanguage used to declare and axiomatize those levels. 
Corresponding to the four metalevels are the four nested metalanguages, 
$\mathtt{meta\mbox{-}lower} \sqsupseteq \mathtt{meta\mbox{-}upper} \sqsupseteq \mathtt{meta\mbox{-}top} \sqsupseteq \mathtt{meta\mbox{-}ur}$,
where each metalanguage axiomatization includes specialization of the one immediately above.
Within each metalevel, the terminology is partitioned into namespaces, 
and various namespaces are collected together into meaningful composites called meta-ontologies. 
The IFF contains meta-ontologies representing category theory, information flow, formal concept analysis, institutions, multitudes, equational logic, first order logic, the common logic standard, etc.
All of the various meta-ontologies in the IFF are anchored to the IFF metastack
$\mathbf{Set} \subseteq \mathbf{Cls} \subseteq \mathbf{Col} \subseteq \mathbf{Ur}$.
The IFF development is largely driven by the principles of \emph{conceptual warrant}, \emph{categorical design} and \emph{institutional logic}.
The main application of the IFF is institutional:
the notion of institutions and their morphisms are being axiomatized in the upper metalevels of the IFF, 
and the lower metalevel of the IFF has axiomatized various institutions
in which semantic integration has a natural expression.
\end{abstract}

\begin{quotation}
\noindent ``\footnotesize{\emph{Philosophy cannot become scientifically healthy without an immense technical vocabulary. We can hardly imagine our great-grandsons turning over the leaves of this dictionary without amusement over the paucity of words with which their grandsires attempted to handle metaphysics and logic. Long before that day, it will have become indispensably requisite, too, that each of these terms should be confined to a single meaning which, however broad, must be free from all vagueness. This will involve a revolution in terminology; for in its present condition a philosophical thought of any precision can seldom be expressed without lengthy explanations.}}'' 
\noindent --- Charles Sanders Peirce, Collected Papers 8:169
\end{quotation}

\section{Introduction}\label{sec-introduction}

Recall reading the classic
\emph{Categories for the Working Mathematician} (\emph{CWM}) \cite{maclane:71}
by Saunders Mac Lane.
This standard reference offers an introduction to the basic concepts of category theory,
such as the categories\footnote{The philosophical notion of category. After all, the term ``category'' in ``category theory'' was purloined from Aristotle and Kant.} ``category'', ``functor'', ``natural transformation'', ``adjunction'', ``monad'' and ``(co)limit'',
and the relationships (some functional) ``object'', ``morphism'' = ``arrow'' (terms are equal when they are synonyms), ``source'', ``target'', ``equivalent to'', ``isomorphic to'', etc.
\emph{CWM} discusses a collection of well-defined concepts such as these, each being represented by a word or phrase in natural language augmented by mathematical notation.
In addition,
\emph{CWM} contains a collection of logical expressions used to represent its meaning.
The logical expressions of \emph{CWM} are built up from its categories and relationships using logical connectives and quantifiers, or are expressed using commutative diagrams.
For example,
consider a page chosen at random from \emph{CWM}, such as page 78.
Some of the categorical concepts encountered on this page include:
``free category'', ``small graph'', ``functor'', ``forgetful functor'', ``$U : \mathbf{Cat} \rightarrow \mathbf{Grph}$'' = ``the forgetful functor from the category of small categories to the category of small graphs'', ``morphism'', ``unique map'', ``category'', ``natural isomorphism'', ``$\mathbf{Set}$'' = ``the category of small sets'', ``function'', ``bijection'', ``natural'', ``isomorphism'', ``adjunction'', ``bifunctor'', ``$X^{\mathrm{op}}{\times}A \rightarrow \mathbf{Set}$'' = ``a $\mathbf{Set}$-valued bifunctor'', ``pair of objects'', ``hom-set'', etc.
Some of the logical expressions encountered on this page include:
``for all modules $A$, $B$, and $C$ over a commutative ring $K$,
there is an isomorphism $\mathrm{Hom}(A\otimes_{K}B, C) \cong \mathrm{Hom}(A, B\otimes_{K}C)$''
and
``for all $A$-morphisms $k : a \rightarrow a^\prime$ and all $X$-morphisms $h : x^\prime \rightarrow x$ the diagrams
\begin{center}
\begin{footnotesize}
\setlength{\unitlength}{0.8pt}
\begin{picture}(337.5,80)(0,0)
\put(0,0){\begin{picture}(125,75)(0,0)
\put(0,50){\makebox(50,25){$A(Fx, a)$}}
\put(0,0){\makebox(50,25){$A(Fx, a^\prime)$}}
\put(75,50){\makebox(50,25){$X(x, Ga)$}}
\put(75,0){\makebox(50,25){$X(x, Ga^\prime)$}}
\put(-30,25){\makebox(50,25){\footnotesize{$A(Fx, k)$}}}
\put(105,25){\makebox(50,25){\footnotesize{$X(x, Gk)$}}}
\put(50,62.5){\makebox(25,25){\footnotesize{$\phi_{x, a}$}}}
\put(50,-12.5){\makebox(25,25){\footnotesize{$\phi_{x, a^\prime}$}}}
\put(53,62.5){\vector(1,0){19}}
\put(53,12.5){\vector(1,0){19}}
\put(25,50){\vector(0,-1){25}}
\put(100,50){\vector(0,-1){25}}
\end{picture}}
\put(200,0){\begin{picture}(125,75)(0,0)
\put(0,50){\makebox(50,25){$A(Fx, a)$}}
\put(0,0){\makebox(50,25){$A(Fx^\prime, a)$}}
\put(75,50){\makebox(50,25){$X(x, Ga)$}}
\put(75,0){\makebox(50,25){$X(x^\prime, Ga)$}}
\put(-30,25){\makebox(50,25){\footnotesize{$A(Fh, a)$}}}
\put(105,25){\makebox(50,25){\footnotesize{$X(h, Ga)$}}}
\put(50,62.5){\makebox(25,25){\footnotesize{$\phi_{x, a}$}}}
\put(50,-12.5){\makebox(25,25){\footnotesize{$\phi_{x^\prime, a}$}}}
\put(53,62.5){\vector(1,0){19}}
\put(53,12.5){\vector(1,0){19}}
\put(25,50){\vector(0,-1){25}}
\put(100,50){\vector(0,-1){25}}
\end{picture}}
\end{picture}
\end{footnotesize}
\end{center}
commute''. 
Some questions immediately arise.
Can \emph{CWM} be formalized, either approximately or fully?
Should it be formalized?
What meaning is captured by formalization?
What meaning is missed?
Should we wait for a more mature foundations of category theory before formalizing \emph{CWM}?
Would a formalization of \emph{CWM} be itself a candidate foundations of category theory?
This paper discusses an effort called the Information Flow Framework (IFF) \cite{iff:homepage},
which can be viewed as an approximate formalization of \emph{CWM},
since a substantial portion of \emph{CWM} has already been axiomatized in the IFF,
but which was originally formulated to be a descriptive category metatheory for object-level ontologies.

\section{Ontologies}\label{sec-ontologies}

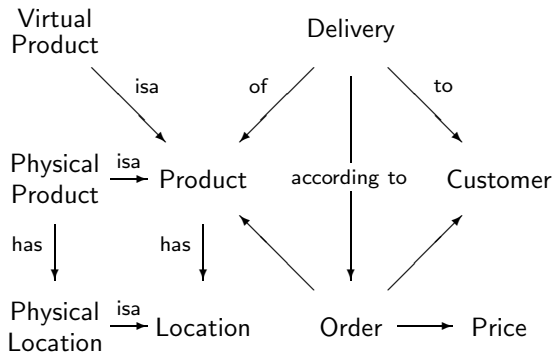
\begin{figure}
\begin{center}
\begin{footnotesize}
\setlength{\unitlength}{0.7pt}
\begin{picture}(320,240)(0,30)
\put(0,160){\makebox(80,80){\shortstack{\sffamily{Virtual}\\\sffamily{Product}}}}
\put(160,160){\makebox(80,80){\sffamily{Delivery}}}
\put(0,80){\makebox(80,80){\shortstack{\sffamily{Physical}\\\sffamily{Product}}}}
\put(80,80){\makebox(80,80){\sffamily{Product}}}
\put(240,80){\makebox(80,80){\sffamily{Customer}}}
\put(0,0){\makebox(80,80){\shortstack{\sffamily{Physical}\\\sffamily{Location}}}}
\put(80,0){\makebox(80,80){\sffamily{Location}}}
\put(160,0){\makebox(80,80){\sffamily{Order}}}
\put(240,0){\makebox(80,80){\sffamily{Price}}}
\put(70,150){\makebox(40,40){\scriptsize{\sffamily{isa}}}}
\put(130,150){\makebox(40,40){\scriptsize{\sffamily{of}}}}
\put(230,150){\makebox(40,40){\scriptsize{\sffamily{to}}}}
\put(60,110){\makebox(40,40){\scriptsize{\sffamily{isa}}}}
\put(5,65){\makebox(40,40){\scriptsize{\sffamily{has}}}}
\put(60,30){\makebox(40,40){\scriptsize{\sffamily{isa}}}}
\put(85,65){\makebox(40,40){\scriptsize{\sffamily{has}}}}
\put(179,100){\makebox(40,40){\scriptsize{\sffamily{according to}}}}
\put(60,180){\vector(1,-1){40}}		
\put(180,180){\vector(-1,-1){40}}	
\put(220,180){\vector(1,-1){40}}	
\put(200,175){\line(0,-1){46}}	
\put(200,113){\vector(0,-1){48}}	
\put(180,60){\vector(-1,1){40}}		
\put(220,60){\vector(1,1){40}}		
\put(40,95){\vector(0,-1){30}}		
\put(120,95){\vector(0,-1){30}}	  
\put(70,120){\vector(1,0){20}}		
\put(70,40){\vector(1,0){20}}		  
\put(225,40){\vector(1,0){30}}		
\end{picture}
\end{footnotesize}
\end{center}
\caption{An e-commerce ontology fragment}
\label{example-ontology}
\end{figure}

Since the IFF is being offered as a metatheory for ontologies,
it is appropriate to ask the question ``What is an ontology?''.
The term ``ontology'' was introduced in 17th century philosophy to represent the study of being,
but had origins in Aristotle's metaphysics.
In the 20th century, 
the term ``ontology'' was borrowed by the artificial intelligence community 
to represent a collection of things that exist.
The meaning has since evolved, and now represents conceptual modeling and knowledge engineering.
The various kinds of ontologies form a spectrum according to formality,
with catalogs and glossaries at one pole, taxonomies in the middle, and formal axiomatic theories at the other pole.
One popular definition is that
``an ontology is a formal, explicit specification of a shared conceptualization. 
It is an abstract model of some phenomena in the world,
explicitly represented as concepts, relationships and constraints,
which is machine-readable and incorporates the consensual knowledge of some community.''
Since it is an abstract model of some phenomena in the world, it is a semantic conception.
Because it uses concepts, relationships and constraints, it is logic-oriented.
Because it is machine-readable, it is formal and explicit.
And since it incorporates the consensual knowledge of some community, it is shared and relative.
Figure~\ref{example-ontology} illustrates a fragment of an E-commerce schema (ontology) taken from the paper\footnote{This paper by C.N.G Dampney and Michael Johnson suggests regarding category theory as a meta-ontology, which fits exactly the spirit of the Information Flow Framework (IFF) project.} \cite{dampney:johnson:01}, 
where concepts are represented as nodes, relationships are represented as edges, and constraints are represented as parallel pairs of edge paths, limits and coproducts.

We distinguish between object-level and metalevel ontologies. 
An object-level ontology represents some aspect of the ``real world''. 
Three levels of object-level ontology have been mentioned.
Lower object-level (domain) ontologies represent the conceptual structures of some community; e.g., physics, chemistry, biology, government, business, etc.
One example of a domain ontology is a physics ontology, consisting of the organization of the community of physicists, a collection of physical subtheories, such as gravity, heat, motion, electricity, magnetism, \dots, and technical terminology.
Another example of a domain ontology is {\em The Gene Ontology (GO)}, an actual bio-ontology for the genetics community, consisting of the organization of the community of geneticists, subtheories, such as molecular-function, biological-process, cellular-component, \ldots, and technical terminology.
Middle object-level ontologies help organized domain ontologies by axiomatizing some abstract facet,
such as the conceptual structures of organizations, the model of an enterprise, the nature of governments, a template for educational institutions, the mathematical structures of events and processes, \ldots.
Upper object-level ontologies are limited to concepts that are meta, generic, abstract and philosophical. 
Upper-level ontologies contain the most general classifications of entities.
Hence, they tend to be domain independent and possess high reusability.
Examples of upper-level ontologies include the WordNet online lexical database, the Standard Upper Ontology (SUO), the Cyc upper ontology, the High Performance Knowledge Base project (HPKB), the Suggested Upper Merged Ontology (SUMO), \ldots.
Of course, there may be overlap between domain and middle-level ontologies, and between middle-level and upper-level ontologies.
And often, upper-level ontologies contain meta-information.

A meta-level ontology is an ontology about ontologies --- it represents some aspect of the organization of object-level ontologies.
In this paper, we identify meta-ontologies and metatheories.
However, object-level ontologies and theories are not identical,
since object-level ontologies can be either populated or not.
We identify unpopulated object-level ontologies with theories [aka schemas],
and we identify populated ontologies with (local) logics [aka databases].
Unpopulated ontologies have only type information.
Populated ontologies have both type and instance information,
plus the classification relationships between these two kinds of things.
This paper is about a metatheory called the Information Flow Framework (IFF),
which is descriptive and category-theory-based.
The fact that the IFF is a descriptive metatheory involves use of the constraint called ``conceptual warrant''.
The fact that the IFF is a category-theory-based metatheory involves use of the motivator called ``the categorical design principle''.

\section{Descriptive Metatheories}\label{sec-descriptive-metatheories}

Distinctions are important in ontology development. 
Two distinctions in particular are important for the IFF:
the monolithic versus modular distinction and the prescriptive versus descriptive 
distinction\footnote{Compare the prescriptive-semantic distinction discussed in the paper \cite{dampney:johnson:01} by Dampney and Johnson.}.
However, these distinctions are sometimes confused
--- although prescriptive approaches are often monolithic and descriptive approaches are often modular,
these two distinctions are conceptually different. 
A monolithic ontology is one-size-fits-all.
The monolithic approach is not compatible with the need for continual revision and consistency checking.
The modular approach, which advocates the lattice and context of theories, is very compatible with these needs. 
The monolithic-modular distinction is important for the maintenance of object-level ontologies (see the discussion below).
However, in this section we are mainly interested in the prescriptive-descriptive distinction,
which is important for meta-level ontologies; that is, metatheories.

All natural languages are inherently dynamic living entities.
However, the dynamic growth of some languages such as English is greatly aided by the way that it is cataloged.
In 1755 the great English lexicographer and literary critic Samuel Johnson published his {\em Dictionary of the English Language}~\cite{johnson:cham}.
This work set the standard for all modern English dictionaries.
However, this standard of lexicography is distinctly different from that used in the related French and Italian languages.
The linguistic purity of French and Italian is to some extent regulated by the Acad\'{e}mie Fran\c{c}aise\footnote{1635--present [\tts{www.academie-francaise.fr}]} and the Accademia della Crusca\footnote{1582--present [\tts{www.accademiadellacrusca.it}]}, respectively.
These two bodies were established ``to prescribe the use of the language''~\cite{winchester:03}.
As the full title\footnote{{\em A Dictionary of the English Language: In Which the Words are Deduced from Their Originals, and Illustrated in Their Different Significations by Examples from the Best Writers}~\cite{johnson:cham}} of Johnson's work implies, 
modern English dictionaries do not {\em prescribe} how the English language should be used, but instead {\em describe} how the language actually is used.
The following quotation from the preface to~\cite{johnson:cham} expands on this descriptive attitude.
\begin{quote}``\footnotesize{\emph{Every increase of knowledge, whether real or fancied, will produce new words, or combination of words. When the mind is unchained from necessity, it will range after convenience; when it is left at large in the fields of speculation, it will shift opinions; as any custom is disused, the words that expressed it must perish with it; as any opinion grows popular, it will innovate speech in the same proportion as it alters practice.}}''
\end{quote}
An ontology is similar to a dictionary, but has greater detail and structure. 
Both dictionaries and ontologies come in two basic philosophies: prescriptive or descriptive.
A descriptive dictionary or ontology describes actual usage.
Most modern dictionaries are descriptive with the Oxford English Dictionary (OED) as a leading example.
As discussed below, the Information Flow Framework (IFF) follows a similar descriptive philosophy.

In this paper we are mainly interested in the analogy between dictionaries and \underline{meta}-ontologies.
First, 
there is a correspondence between the \emph{builders} of dictionaries and those of meta-ontologies.
Corresponding to the lexicographers\footnote{Compare Samuel Johnson's definition~\cite{johnson:cham} of a lexicographer as a ``a writer of dictionaries, a harmless drudge, that busies himself in tracing the original, and detailing the signification of words''.}, 
such as Samuel Johnson or James Murray of the OED, who create dictionaries, 
are the ontologicians (mathematicians, particularly category-theorists) who create meta-ontologies.
Second, 
there is a correspondence between the \emph{source material} used by dictionaries and that used by meta-ontologies.
As discussed in \cite{winchester:03}, the entries placed and described in dictionaries have three sources: 
(i) terms borrowed from other dictionaries, (ii) new terms used to express concepts in works of literature, and (iii) new terms used to express concepts in everyday speech.
By analogy, the concepts axiomatized in meta-ontologies originate from three sources: 
(i) terms borrowed from other meta-ontologies, (ii) new metalevel terms used to express concepts in object-level ontologies, and (iii) new metalevel terms used to express the conceptual structure of a community.
For both dictionaries and ontologies, the second source is most important.
Lexicographers use works of literature as the main source for dictionaries,
whereas ontologicians use the meta, generic and abstract terminology of object-level ontologies as the main source for meta-ontologies.
For meta-ontologies,
this constraining process is known as ``conceptual warrant'' and is discussed in more detail below.
Third, 
there is a correspondence between the \emph{source creators} used by dictionaries and those used by meta-ontologies.
Corresponding to the literary figures who originate new terms in works of literature are 
the knowledge engineers who originate and use new metalevel terms in object-level ontologies.

\section{The Standard Upper Ontology (SUO)}\label{sec-suo}

The IEEE P1600.1 Standard Upper Ontology (SUO) project \cite{suo:homepage},
which operates under the auspices of the IEEE Standards Association (IEEE-SA),
aims to specify an upper ontology that will provide a structure and a set of general concepts upon which object-level ontologies can be constructed.
According to the IEEE-SA, ``a standard is a published document that sets out specifications and procedures designed to ensure that a material, product, method, or service meets its purpose and consistently performs to its intended use''\footnote{According to the dictionary \cite{m-w:col11},
a standard has the sense of
``something established by authority, custom, or general consent as a model or example'',
``something set up and established by authority as a rule for the measure of quantity, weight, extent, value, or quality'',
or ``a structure built for or serving as a base or support''.
A standard ``applies to any definite rule, principle, or measure established by authority''.}.
It is also understood that such a standard may involve some agreement about the conformance to implementations of the standard.
As understood by the SUO working group, an upper ontology is limited to concepts that are meta, generic, abstract and philosophical.
It is anticipated that the SUO will be useful in data interoperability, information search and retrieval, automated inferencing, and natural language processing.
The SUO follows Robert's Rules of Order
--- there are various discussions on the email list, after which the chair can possibly call for a vote.
The SUO has been active for about five years.
During this time, it has approved several resolutions.
In chronological order, these are described as follows. 
\begin{description}
\item[2001 August:] Information Flow Framework (IFF). 
The SUO IFF project\footnote{[\tts{suo.ieee.org/IFF/}]} was the first proposal passed by the SUO working group.
It is a descriptive category metatheory that represents the structural aspect of the SUO.
The IFF is involved with concepts that are meta\footnote{According to the dictionary \cite{m-w:col11},
the term `meta-' is a prefix meaning
a ``more highly organized or more comprehensive form of'' something,
and is ``used with the name of a discipline to designate a new but related discipline designed to deal critically with the original one''.
Examples include:
a {\em metatheory} is a theory whose subject matter is another theory;
a {\em metalanguage} is a language used to talk about other languages; and
{\em metamathematics} is the field of study concerned with the formal structure and properties of mathematical systems.}, generic and abstract.
The philosophy, approach, architecture and methods of the IFF are described in this paper
(see
Sections~\ref{sec-introduction}, \ref{sec-ontologies} and \ref{sec-descriptive-metatheories} for the philosophy,
Sections~\ref{sec-design-guidelines} and \ref{sec-development-phases} for the approach,
Sections~\ref{sec-architecture}, \ref{sec-metastack}, \ref{sec-lower-metalevel} and \ref{sec-metalanguages} for the architecture and
Section~\ref{sec-application} for the methods).
As described in this paper, the IFF might be called the Standard Meta Ontology (SMO).
In its applicational aspect,
the IFF project uses techniques and concepts from the fields of information flow (\cite{barwise:seligman:97}), formal concept analysis (\cite{ganter:wille:99}), category theory (\cite{maclane:71}) and institutions (\cite{goguen:burstall:92}).
\item[2003 June:] Suggested Upper Merged Ontology (SUMO). 
The SUO SUMO project\footnote{[\tts{suo.ieee.org/SUO/SUMO/}]} is sponsored and developed by the Teknowledge Corporation\footnote{[\tts{www.teknowledge.com}]}.
The SUMO proposal initially failed its vote, principally due to advocacy of a monolithic philosophy for ontology development. 
About this time, there was a vigorous debate on the SUO email list about the monolithic--modular distinction (See Section~\ref{sec-application} for a discussion about modularity.).
The modular philosophy for ontology development is advocated by the SUO Lattice of Theories project.
With passage of this proposal, there is an expressed intent that SUO SUMO working group will collaborate with the SUO Lattice of Theories working group.
\item[2003 June:] Lattice of Theories (LOT). 
The SUO Lattice of Theories project intends to develop a standard for ontology specification and registration.
The standard will be based on the contributions of other SUO candidate projects.
The standard will specify an ontology registry, 
such as the metadata registries specified by the International Organization for Standardization (ISO) standard ISO/IEC IS 11179-3, 
but with extensions that are needed in order to define and relate ontologies.
The ontology registry shall be organized as a collection of modules, related in a generalization/specialization hierarchy.
Each module shall consist of a theory together with documentation and other metadata. 
The theory shall consist of axioms and definitions stated in a logic-based language, such as those in the Common Logic (CL) framework.
The standard shall include the specification of a methodology for testing the theory part of any module for consistency, relating theories to one another in the generalization/specialization hierarchy, and combining two or more theories to derive a new theory that is larger and more specialized than the theories from which it was derived (See Section~\ref{sec-application} for a discussion and mathematical formulation of the ``lattice of theories'' construction.). 
\item[2003 October:] OpenCyc Ontology. 
The OpenCyc ontology\footnote{[\tts{www.opencyc.org}]},
which contains about 5,000 concepts and 50,000 axioms (aka rules),
is the open source version of the Cyc ontology.
The Cyc ontology,
which contains about 300,000 concepts and 3 million axioms,
is a large and general knowledge base whose intended use is for commonsense reasoning. 
It has spent over 600 person-years in a development effort over the past 17 years. 
The SUO working group would like to dismantle the OpenCyc ontology into meaningful components and reassemble them within the flexible and dynamically changeable structure of the ``lattice of theories'' framework (Again, see Section~\ref{sec-application} for a discussion and mathematical formulation of the ``lattice of theories'' construction.).
\item[2003 October:] SUO 4D Ontology.
The SUO 4D Ontology project\footnote{[\tts{suo.ieee.org/SUO/SUO-4D/}]} intends to develop an ontology based on the 4-dimensional paradigm.
A starting point for the development of this ontology is contained in the ISO standard ISO/FDIS 15926-2, 
a lifecycle integration of process plant data including oil and gas production facilities.  
The approach will be to develop the ontology as reusable components within the institutional aspect of the IFF and to develop mappings to other ontologies within this framework. 
\item[2004 May:] Multi-Source Ontology (MSO).
The SUO MSO is based at the WebKB-2 knowledge server\footnote{[\tts{www.webkb.org}]}, a knowledge server that permits Web users to browse and update private knowledge bases on their machines and a large shared knowledge base on the server machine. 
The ontology of the shared knowledge base is currently an integration of various top-level ontologies (e.g. Sowa, Dolce, the Lifecycle Integration Schema, the Natural Semantic Metalanguage, OWL, DAML+OIL, KIF and the Dublin Core) and a lexical ontology derived from an extension and correction of the noun-related part of WordNet 1.7. 
\end{description}

\section{The IFF Design Guidelines}\label{sec-design-guidelines}

\begin{figure}
\begin{center}
\setlength{\unitlength}{0.75pt}
\begin{picture}(300,250)(0,0)
\put(0,0){\line(1,2){125}}
\put(300,0){\line(-1,2){125}}
\put(125,250){\line(1,0){50}}
\put(25,51){\line(1,0){250}}
\put(25,49){\line(1,0){250}}
\put(0,0){\line(1,0){300}}
\thinlines
\put(100,200){\line(1,0){100}}
\put(75,150){\line(1,0){150}}
\put(50,100){\line(1,0){200}}
\put(125,50){\begin{picture}(50,200)(0,0)
\put(2,2){\line(0,1){196}}
\put(48,162){\line(0,1){36}}
\put(48,2){\line(0,1){150}}
\put(2,2){\line(1,0){46}}
\put(2,198){\line(1,0){46}}
\put(0,1){\line(0,1){199}}
\put(50,162){\line(0,1){38}}
\put(50,1){\line(0,1){151}}
\put(0,150){\makebox(50,50){\Huge{$\mathsf{C}$}}}
\put(0,100){\makebox(50,50){\Huge{$\mathsf{O}$}}}
\put(0,50){\makebox(50,50){\Huge{$\mathsf{R}$}}}
\put(0,0){\makebox(50,50){\Huge{$\mathsf{E}$}}}
\end{picture}}
\put(65,53){\begin{picture}(50,200)(0,0)
\put(75,150){\makebox(50,10)[br]{\scriptsize{\emph{generic}}}}
\put(95,110){\makebox(50,10)[br]{\scriptsize{\emph{very}}}}
\put(100,100){\makebox(50,10)[br]{\scriptsize{\emph{large}}}}
\put(125,50){\makebox(50,10)[br]{\scriptsize{\emph{large}}}}
\put(150,1){\makebox(50,10)[br]{\scriptsize{\emph{small}}}}
\end{picture}}
\put(-115,0){\begin{picture}(100,250)(0,0)
\put(100,52){\line(0,1){196}}
\put(100,248){\line(1,0){5}}
\put(100,52){\line(1,0){5}}
\put(100,2){\line(0,1){46}}
\put(100,48){\line(1,0){5}}
\put(100,2){\line(1,0){5}}
\put(-50,50){\makebox(140,200)[r]{\large{\textsl{metalevel}}}}
\put(-50,0){\makebox(140,50)[r]{\large{\textsl{object level}}}}
\end{picture}}
\end{picture}
\end{center}
\caption{The IFF Architecture (iconic version)}
\label{architecture-icon}
\end{figure}
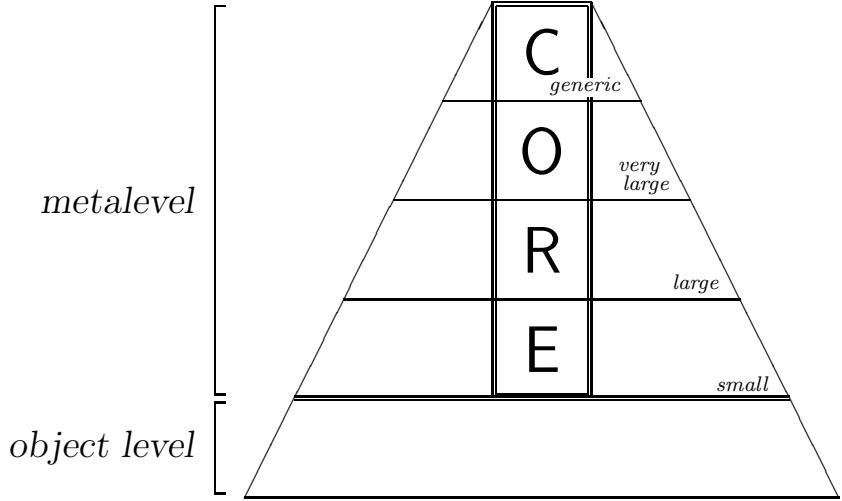

The Information Flow Framework (IFF) (Figure~\ref{architecture-icon}) is a descriptive category metatheory that is intended to represent the structural aspect of the SUO.
In the development of the IFF, certain guidelines have proven to be very important.
These are all predicated on the goal of building a category-theoretic metatheory for object-level ontologies.
Initially,
this metatheory was designed to represent first order logic,
its languages, theories, model-theoretic structures and (local) logics,
including satisfaction and fibrations based at languages.
Later, this metatheory incorporated the theory of institutions of Goguen and Burstall.
The most important guideline in the development of the IFF --- what we might call the meta-guideline --- is to follow the intuitions of the working category-theorist.
Such intuitions represent naive category theory\footnote{The meaning of naive here is not pejorative.
It means \cite{m-w:col11} primitive, natural, intuitive, first-formed, primary, not evolved or elemental.}.
In practice, we initially formulate any IFF module as a set-theoretic axiomatization using a first order expression.
Then,
by eliminating quantifiers and logical connectives,
we attempt to move, morph or transform this set-theoretic axiomatization toward a category-theoretic axiomatization.
A convenient rule-of-thumb is to ``keep it simple''.
From the foundational standpoint, this means that we start with no assumptions at all.
However, 
in view of the Cantor diagonal argument, 
as a first step we assume a slender hierarchy called the IFF metastack.

The IFF metastack, 
which is the core hierarchy of the IFF, is represented as the central axis or spindle in Figure~\ref{architecture-icon}.
The content of the metastack is the axiomatization for sets and functions (and for convenience, binary relations) in four core modules at four different set-theoretic levels, 
the ``small'' in the lower metalevel, 
the ``large'' in the upper metalevel, 
the ``very large'' in the top metalevel and 
the ``generic'' in the ur metalevel. 
Currently, the axiomatization for the metastack has a ``conical'' shape\footnote{The conical shape of this axiomatization seems natural so far, even with the axiomatization for 2-categories. However, with the proper placement of the axiomatization for institutions in the upper metalevel, any strong appeal for application of the categorical design principle to this institutional axiomatization would call for exponents in the top metalevel. This would give the metastack axiomatization more of a ``cylindrical'' shape.}, with
an axiomatization for categories introduced in the ur metalevel, 
an axiomatization for finite limits introduced in the top metalevel, 
an axiomatization for exponents and finite colimits introduced in the upper metalevel, and
an axiomatization for subobject classifiers\footnote{This axiomatization has not been realized so far, since it is not needed, and hence, at this time would violate the principle of conceptual warrant (see below). However, we anticipate its need.}
and general limits/colimits introduced in the lower metalevel\footnote{Assuming a conical shape,
an {\em abstract metastack} 
$\mathbf{\mathsf{M}} 
= \langle \mathbf{\mathsf{M_4}} \supseteq \mathbf{\mathsf{M_3}} \supseteq \mathbf{\mathsf{M_2}} \supseteq \mathbf{\mathsf{M_1}} \rangle$
is define to be a category $\mathbf{\mathsf{M_4}}$ at the $\mbox{4}^{\mathrm{th}}$ metalevel (Ur)
restricted to a finitely complete category $\mathbf{\mathsf{M_3}}$ at the $\mbox{3}^{\mathrm{rd}}$ metalevel (Top) 
restricted to a cartesian closed category $\mathbf{\mathsf{M_2}}$ at the $\mbox{2}^{\mathrm{nd}}$ metalevel (Upper)
restricted to a topos $\mathbf{\mathsf{M_1}}$ at the $\mbox{1}^{\mathrm{st}}$ metalevel (Lower).}.
This effectively distributes topos axioms over the four metalevels.
The categories $\mathbf{Set} \subseteq \mathbf{Cls} \subseteq \mathbf{Col} \subseteq \mathbf{Ur}$
in the IFF metastack, which correspond to the four metalevels mentioned above, are axiomatized in the lower core (\tts{IFF-LCO}), the upper core (\tts{IFF-UCO}), the top core (\tts{IFF-TCO}) and the ur (\tts{IFF-UR}) meta-ontologies, respectively.
Here, 
all inclusions preserve composition and identities (are functorial),
the first two preserve finite limits, and 
the first preserves the Cartesian closed structure\footnote{Assuming a conical shape,
a {\em morphism of abstract metastacks}
$\mathbf{\mathsf{H}} : \mathbf{\mathsf{M}} \rightarrow \mathbf{\mathsf{M'}}$
consists of 
a functor (morphism of categories) 
$\mathbf{\mathsf{H_4}} : \mathbf{\mathsf{M_4}} \rightarrow \mathbf{\mathsf{M_4}}$
at the $\mbox{4}^{\mathrm{th}}$ metalevel,
restricted to a morphism of finitely complete categories 
$\mathbf{\mathsf{H_3}} : \mathbf{\mathsf{M_3}} \rightarrow \mathbf{\mathsf{M_3}}$
at the $\mbox{3}^{\mathrm{rd}}$ metalevel,
restricted to a morphism of cartesian closed categories
$\mathbf{\mathsf{H_2}} : \mathbf{\mathsf{M_2}} \rightarrow \mathbf{\mathsf{M_2}}$
at the $\mbox{2}^{\mathrm{nd}}$ metalevel,
restricted to a (geometric) morphism of toposes
$\mathbf{\mathsf{H_1}} : \mathbf{\mathsf{M_1}} \rightarrow \mathbf{\mathsf{M_1}}$
at the $\mbox{1}^{\mathrm{st}}$ metalevel.}.

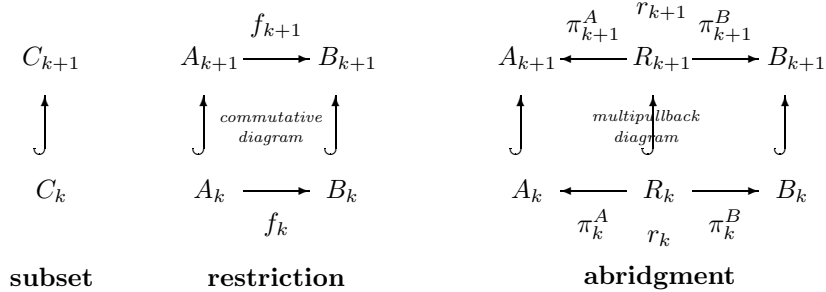
\begin{figure}
\begin{center}
\begin{sffamily}
\begin{footnotesize}
\begin{picture}(280,115)(-14,-20)
\put(-20,0){\begin{picture}(25,75)(0,0)
\put(0,50){\makebox(25,25){$C_{k+1}$}}
\put(0,0){\makebox(25,25){$C_{k}$}}
\put(10,28.5){\begin{picture}(0,20)(0,0)
\put(0,0){\vector(0,1){20}}
\put(-2,0){\oval(4,4)[br]}
\put(-2,0){\oval(4,4)[bl]}
\end{picture}}
\put(0,-30){\makebox(25,20){\rmfamily{\bfseries{subset}}}}
\end{picture}}
\put(40,0){\begin{picture}(75,75)(0,0)
\put(0,50){\makebox(25,25){$A_{k+1}$}}
\put(0,0){\makebox(25,25){$A_{k}$}}
\put(52,50){\makebox(25,25){$B_{k+1}$}}
\put(50,0){\makebox(25,25){$B_{k}$}}
\put(25,62.5){\makebox(25,25){$f_{k+1}$}}
\put(25,-12.5){\makebox(25,25){$f_{k}$}}
\put(10,28.5){\begin{picture}(0,20)(0,0)
\put(0,0){\vector(0,1){20}}
\put(-2,0){\oval(4,4)[br]}
\put(-2,0){\oval(4,4)[bl]}
\end{picture}}
\put(60,28.5){\begin{picture}(0,20)(0,0)
\put(0,0){\vector(0,1){20}}
\put(-2,0){\oval(4,4)[br]}
\put(-2,0){\oval(4,4)[bl]}
\end{picture}}
\put(25,62.5){\vector(1,0){25}}
\put(25,12.5){\vector(1,0){25}}
\put(0,-30){\makebox(75,20){\rmfamily{\bfseries{restriction}}}}
\put(-2,28){\makebox(75,25){\rmfamily{\emph{\tiny{commutative}}}}}
\put(-2,20){\makebox(75,25){\rmfamily{\emph{\tiny{diagram}}}}}
\end{picture}}
\put(160,0){\begin{picture}(125,75)(0,0)
\put(0,50){\makebox(25,25){$A_{k+1}$}}
\put(51,68){\makebox(25,25){$r_{k+1}$}}
\put(51,50){\makebox(25,25){$R_{k+1}$}}
\put(100,0){\makebox(25,25){$B_{k}$}}
\put(0,0){\makebox(25,25){$A_{k}$}}
\put(50,0){\makebox(25,25){$R_{k}$}}
\put(50,-18){\makebox(25,25){$r_{k}$}}
\put(102,50){\makebox(25,25){$B_{k+1}$}}
\put(25,62.5){\makebox(25,25){$\pi^{A}_{k+1}$}}
\put(75,62.5){\makebox(25,25){$\pi^{B}_{k+1}$}}
\put(25,-12.5){\makebox(25,25){$\pi^{A}_{k}$}}
\put(75,-12.5){\makebox(25,25){$\pi^{B}_{k}$}}
\put(10,28.5){\begin{picture}(0,20)(0,0)
\put(0,0){\vector(0,1){20}}
\put(-2,0){\oval(4,4)[br]}
\put(-2,0){\oval(4,4)[bl]}
\end{picture}}
\put(60,28.5){\begin{picture}(0,20)(0,0)
\put(0,0){\vector(0,1){20}}
\put(-2,0){\oval(4,4)[br]}
\put(-2,0){\oval(4,4)[bl]}
\end{picture}}
\put(110,28.5){\begin{picture}(0,20)(0,0)
\put(0,0){\vector(0,1){20}}
\put(-2,0){\oval(4,4)[br]}
\put(-2,0){\oval(4,4)[bl]}
\end{picture}}
\put(50,62.5){\vector(-1,0){25}}
\put(50,12.5){\vector(-1,0){25}}
\put(75,62.5){\vector(1,0){25}}
\put(75,12.5){\vector(1,0){25}}
\put(25,-30){\makebox(75,20){\rmfamily{\bfseries{abridgment}}}}
\put(20,28){\makebox(75,25){\rmfamily{\emph{\tiny{multipullback}}}}}
\put(20,20){\makebox(75,25){\rmfamily{\emph{\tiny{diagram}}}}}
\end{picture}}
\end{picture}
\end{footnotesize}
\end{sffamily}
\end{center}
\caption{The three fundamental generic relations}
\label{subset-restriction-abridgment}
\end{figure}

These preservation properties represent the fact that the axiomatization at each metalevel includes the specialization of the axiomatization at the next higher metalevel.
Specialization is a technical term here,
which means the exact adaptation of the core terminology and axiomatization at one metalevel to the next lower metalevel by use of the three fundamental generic relations of subset, (function) restriction and (binary relation) abridgment that are diagrammatically described in Figure~\ref{subset-restriction-abridgment},
where $k$ is a particular metalevel, $k{+}1$ is the next higher metalevel, the set $C_{k}$ is a subset of the set $C_{k+1}$, the function $f_{k}$ is a restriction of the function $f_{k+1}$, and the relation $r_{k}$ with extent $R_{k}$ is an abridgment of the relation $r_{k+1}$ with extent $R_{k+1}$.
Table~\ref{metastack-code} gives a brief glance at the IFF code that represents specialization of the core set notions.

\begin{table}
\begin{center}
\begin{scriptsize}
\begin{tabular}{|p{4in}|} \hline
\\ 
\ttbf{IFF-UR}: ``\emph{An Ur-object represents the notion of a generic set.}''
\begin{verbatim}
(forall (?x (object ?x))
    (thing ?x))
\end{verbatim} \\ \hline\hline
\\
\ttbf{IFF-TCO}: ``\emph{A collection represents the notion of a very large set. There is an Ur-object of all collections. Any collection is a UR-object. The Ur-object of all collections is not itself a collection.}''
\begin{verbatim}
(ur:object collection) 
(forall (?c (collection ?c))
    (ur:object ?c)) 
(not (collection collection))
\end{verbatim}
\\ \hline\hline
\\
\ttbf{IFF-UCO}: ``\emph{A class represents the notion of a large set. There is a collection of all classes. Classes are mainly used in IFF to specify the object and morphism collections of large categories such as the category of all small sets and functions} \rmb{Set} \emph{and the category of all small classifications and infomorphisms} \rmb{Clsn}. \emph{Every class is a collection. The collection of all classes is not itself a class.}''
\begin{verbatim}
(vlrg.set:collection class) 
(ur:subobject class vlrg.set:collection) 
(not (class class))
\end{verbatim}
\\ \hline
\end{tabular}
\end{scriptsize}
\end{center}
\caption{Code for the IFF Metastack}
\label{metastack-code}
\end{table}

Another guideline is that quantification should be over a specific collection.
This capability is realized in the IFF \tts{metashell} (see Section~\ref{sec-metalanguages} on metalanguages),
and hence all metalanguages,
by the abbreviation called restricted quantification.
A third guideline in the IFF development,
is to use functions wherever possible.
This encourages the elimination of quantifiers.
A final guideline is that as a general rule, universes are kept away from power operators (and hence also exponent operators).
This is part of the injunction to avoid contradictions (paradoxes) due to Cantor (and Russell etc.).
In practice, we have little or no need for universes 
--- we use the term ``thing'' in the Ur and Top core ontologies, but this has minimal use.

\section{The IFF Development Phases}\label{sec-development-phases}

The IFF has gone through two phases of development.
The IFF Category Theory (meta) Ontology (\tts{IFF-CAT}) is distinguished by being the first meta-ontology axiomatized during the first phase of development.
It indicated necessary ingredients in the core axiomatization now known as the metastack.
The \tts{IFF-CAT} has axiomatizations for large categories, large functors, large natural transformations, large adjunctions, large monads, and limits/colimits.
Perhaps \tts{IFF-CAT} is in some sense an entry-level axiomatization for the category of all categories.
A non-starter for the remaining part of the first phase was a topos axiomatization.
This received objections from the SUO working group, in part due to its lack of support by motivating examples.
Rejection of the topos axiomatization prompted the idea of conceptual warrant.
The second phase of the IFF was designed bottom-up and followed conceptual warrant as its guideline.
The most important metalogic terminology incorporated in the second phase references the various concepts related to finite limits.
For example,
the desire to axiomatize composition of class functions in the upper metalevel requires the pullback concept at the top metalevel.
The IFF is now well within its third phase of development,
which involves conversion of the truth construction to the theory of institutions
and reorganization of the metastack using the ``adjunctive axiomatization'' technique illustrated in Lawvere and Rosebrugh \cite{lawvere:rosebrugh:03}.
A major goal for the third phase is to complete the categorical design principle at the lower metalevel and to initiate it at the upper metalevel.
In order to accomplish this,
we have extended composition to all levels of the metastack and 
we have introduced the important exponent metalogic concept in the upper core.
A fourth phase is envisioned in the future, 
when the concepts of fibrations and indexed categories will be axiomatized.
These are important for the axiomatizations of institutions and fibring logics.

During the IFF development,
four concepts have eventually emerged as important.
One of these concepts, the IFF metastack, which originated in phase three of IFF development, was discussed above.
The other three concepts are principles for IFF development.
In chronological order these are
(1) the principle of conceptual warrant,
(2) the principle of categorical design and
(3) the principle of institutional logic.
Conceptual warrant restricts the introduction of upper metalevel terminology,
whereas categorical design forces the introduction of this terminology,
principally in the core.

\begin{quotation}
\noindent {\bfseries Principle: conceptual warrant.} All IFF terminology should require conceptual warrant for their existence:
any term that appears in (and is axiomatized by) a metalanguage should reference a concept needed in a lower metalevel or object level axiomatization.
\end{quotation}
The principle of conceptual warrant originated in phase one of IFF development.
The terminology appearing in any standardization meta-ontology will exert authority.
Because of this, in selecting which terminology to specify in the IFF,
we utilize the notion of ``conceptual warrant''.
Warrant means evidence for, or token of, authorization. 
Conceptual warrant is an adaptation of the librarianship notion of literary warrant.
According to the Library of Congress, its collections serve as the literary warrant 
(i.e., the literature on which the controlled vocabulary is based) 
for the Library of Congress subject headings system.
In the same fashion,
the object-level and lower metalevel terminology of the IFF serves as the conceptual warrant for the IFF upper metalevel axiomatization.

\begin{quotation}
\noindent {\bfseries Principle: categorical design.} The design of a module at any particular metalevel should adhere to the property that its axiomatic representation is strictly category-theoretic:
All axioms use terms from the metalanguage at that metalevel.
No axioms use explicit logical notation:
No variables, quantification (`$\forall$', `$\exists$') or logical connectives (`$\wedge$', `$\vee$', `$\neg$', `$\Rightarrow$',`$\Leftrightarrow$') are used.
\end{quotation}
The principle of categorical design originated in phase two of IFF development.
The goal of this principle has been to simplify the IFF axiomatization 
--- first order expression would be reduced to term-rewriting.
The peripheral (non-core) modules in the lower IFF metalevel have the tripartite form:
outer category namespace, inner object and morphism namespaces (see Section~\ref{sec-lower-metalevel} for more discussion on this and Figure~\ref{lower-core-format} for a visualization). 
The outer namespace fully conforms to the categorical design principle.
The inner namespaces conform to it to a great extent (80--90\%).
The categorical design principle was originally expressed for the lower metalevel (``the small''), but with composition extended upward this principle also seems appropriate for the upper metalevel (``the large''); in particular, it seems appropriate for axiomatizing the IFF Category Theory (meta) Ontology (\tts{IFF-CAT}).
For example, Table~\ref{category-code} contains code for the concept of a (large) category,
which fully conforms to the categorical design principle.

\begin{table}
\begin{center}
\begin{scriptsize}
\begin{tabular}{|p{4in}|} \hline
\\ 
\ttbf{IFF-CAT}: ``\emph{A (large) category can be thought of as a special kind of graph --- a graph with monoidal properties.
It consists of an underlying graph, a composition graph morphism and an identity graph morphism, both with an identity object function.}''
\begin{verbatim}
(vlrg.set:collection category)
(vlrg.ftn:function graph) 
(vlrg.ftn:function underlying) (= underlying graph) 
(= (vlrg.ftn:source graph) category)
(= (vlrg.ftn:target graph) lrg.gph.obj:graph)
(vlrg.ftn:function graph-pair)
(= (vlrg.ftn:source graph-pair) category)
(= (vlrg.ftn:target graph-pair) lrg.gph.obj:multipliable-pair)
(= graph-pair (vlrg.lim.pbk.obj:pairing [graph graph]))
(vlrg.ftn:function mu)
(= (vlrg.ftn:source mu) category)
(= (vlrg.ftn:target mu) lrg.gph.mor:2-cell)
(= (vlrg.ftn:composition [mu lrg.gph.mor:source])
   (vlrg.ftn:composition [graph-pair lrg.gph.obj:multiplication]))
(= (vlrg.ftn:composition [mu lrg.gph.mor:target]) graph)
(vlrg.ftn:function eta)
(= (vlrg.ftn:source eta) category)
(= (vlrg.ftn:target eta) lrg.gph.mor:2-cell)
(= (vlrg.ftn:composition [eta lrg.gph.mor:source])
   (vlrg.ftn:composition (vlrg.ftn:composition
       [graph lrg.gph.obj:object]) lrg.gph.obj:unit]))
(= (vlrg.ftn:composition [eta lrg.gph.mor:target]) graph)
\end{verbatim}
\\ \hline
\end{tabular}
\end{scriptsize}
\end{center}
\caption{Code for (large) categories}
\label{category-code}
\end{table}

\begin{quotation}
\noindent {\bfseries Principle: institutional logic\footnote{Suggested to the author by Joseph Goguen (personal communication).}.} All logics used in the IFF application (see Section~\ref{sec-application}) should be formulated as institutions.
\end{quotation}
The principle of institutional logic originated in phase three of IFF development.
The theories of information flow and formal concept analysis
(and hence, effectively the theory of institutions) 
have been used throughout the IFF development.
This use has centered on the ``truth construction''.
The truth construction is institutional \cite{kent:dagstuhl},
consisting of a classification functor
$\mathsf{clsn} : \mathsf{Sign} \rightarrow \mathsf{CLSN}$
or a concept lattice functor
$\mathsf{clg} : \mathsf{Sign} \rightarrow \mathsf{CLG}$,
where 
$\mathsf{Sign}$ is the category of signatures,
and $\mathsf{CLSN}$ and $\mathsf{CLG}$ are the equivalent categories \cite{kent:02} of (large) classifications and (large) concept lattices, respectively.
The truth classification was discussed in Barwise and Seligman \cite{barwise:seligman:97}\footnote{See example 4.6 on page 71.}.
In the IFF development,
the truth concept lattice,
which was defined to be the concept lattice of the truth classification,
its equivalent intensional aspect of closed theories,
or its equivalent construction of theories under entailment order,
was promoted by the IFF as the proper representation of the ``lattice of theories'' notion advocated by the SUO working group.
As noted in \cite{goguen:logic},
truth is not dyadic between models and sentences,
but triadic between models, sentences and signatures.
This corresponds to the contextual dependency in the semiotics of Charles Sanders Peirce,
which uses signs, objects and interpretants.
In formal concept analysis triadic concept lattices \cite{lehmann:wille} have been used to formalize this.
However,
the representation of contexts as the logics of institutions has greater advantages.

\section{The IFF Architecture}\label{sec-architecture}

\begin{sloppypar}
The IFF architecture (Figure~\ref{architecture-details}) consists of metalevels, namespaces and meta-ontologies.
Within each level, the terminology is partitioned into namespaces. 
The number of namespaces and the content may vary over time: 
new namespaces may be created or old namespaces may be deprecated, 
and new terminology and axiomatization within any particular namespace may change (new versions).
In addition, within each level, various namespaces are collected together into meaningful composites called meta-ontologies. 
At any particular metalevel, 
these meta-ontologies cover all the namespaces at that level, but they may overlap. 
The number of meta-ontologies and the content of any meta-ontology may vary over time: 
new meta-ontologies may be created or old meta-ontologies may be deprecated, 
and new namespaces within any particular meta-ontology may change (new versions).
Table~\ref{meta-ontologies} presents a list of IFF meta-ontologies organized by metalevel.
All URLs in this table should be prefixed with the string `\ttf{http://suo.ieee.org/IFF/}'.
\end{sloppypar}

\begin{figure}
\begin{center}
\begin{sffamily}
\begin{footnotesize}
\begin{picture}(350,130)(0,-15)
\linethickness{0.2pt}
\put(-80,37.5){\makebox(60,25)[r]{\normalsize{\textsl{metalevel}}}}
\put(-80,-27){\makebox(60,25)[r]{\normalsize{\textsl{object level}}}}
\put(-13,2){\line(0,1){96}}
\put(-13,98){\line(1,0){3}}
\put(-13,2){\line(1,0){3}}
\put(-13,-4){\line(0,-1){21}}
\put(-13,-4){\line(1,0){3}}
\put(-13,-25){\line(1,0){3}}
\put(42,2){\framebox(26,96){}}
\put(0,100){\begin{picture}(350,25)(0,0)
\put(40,0){\makebox(30,25){\tt{set}}}
\put(70,0){\makebox(45,25){\tt{cat}}}
\put(115,0){\makebox(35,25){\tt{gph}}}
\put(150,0){\makebox(50,25){\tt{dbl-cat}}}
\put(200,0){\makebox(50,25){\tt{2-cat}}}
\put(250,0){\makebox(50,25){\tt{ins}}}
\put(300,0){\makebox(50,25){\tt{fol}}}
\end{picture}}
\put(0,75){\begin{picture}(70,25)(0,0)
\put(0,0){\makebox(40,25)[r]{\tt{ur = 4}  }}
\put(40,0){\framebox(30,25){core}}
\end{picture}}
\put(0,50){\begin{picture}(250,25)(0,0)
\put(0,0){\makebox(40,25)[r]{\tt{vlrg = 3}  }}
\put(40,0){\framebox(30,25){core}}
\put(70,0){\framebox(45,25){category}}
\put(115,0){\framebox(35,25){graph}}
\put(150,0){\begin{picture}(50,25)(0,0)
\put(0,0){\framebox(50,25){}}\put(0,12.5){\makebox(50,12.5){\underline{double}}}\put(0,0){\makebox(50,12.5){\underline{category}}}
\end{picture}}
\put(200,0){\framebox(50,25){\underline{2-category}}}
\end{picture}}
\put(0,25){\begin{picture}(300,25)(0,0)
\put(0,0){\makebox(40,25)[r]{\tt{lrg = 2}  }}
\put(40,0){\framebox(30,25){core}}
\put(70,0){\framebox(45,25){\underline{category}}}
\put(115,0){\framebox(35,25){graph}}
\put(150,0){\begin{picture}(50,25)(0,0)
\put(0,0){\framebox(50,25){}}\put(0,12.5){\makebox(50,12.5){double}}\put(0,0){\makebox(50,12.5){category}}
\end{picture}}
\put(200,0){\framebox(50,25){2-category}}
\put(250,0){\framebox(50,25){\underline{institution}}}
\end{picture}}
\put(0,0){\begin{picture}(350,25)(0,0)
\put(0,0){\makebox(40,25)[r]{\tt{sml = 1}  }}
\put(40,0){\framebox(30,25){core}}
\put(70,0){\framebox(45,25){category}}
\put(115,0){\framebox(35,25){\underline{graph}}}
\put(150,0){\begin{picture}(50,25)(0,0)
\put(0,0){\framebox(50,25){}}\put(0,12.5){\makebox(50,12.5){double}}\put(0,0){\makebox(50,12.5){category}}
\end{picture}}
\put(200,0){\framebox(50,25){2-category}}
\put(250,0){\framebox(50,25){institution}}
\put(350,0){\makebox(35,25){\Large{\ldots}}}
\put(300,0){\begin{picture}(50,25)(0,0)
\put(0,0){\framebox(50,25){}}\put(0,12.5){\makebox(50,12.5){\underline{first order}}}\put(0,0){\makebox(50,12.5){\underline{logic}}}
\end{picture}}
\end{picture}}
\put(0,-27){\begin{picture}(350,25)(0,0)
\put(0,0){\makebox(40,25)[r]{\tt{obj = 0}  }}
\put(40,0){\framebox(310,25){}}
\put(40,12.5){\framebox(25,12.5){SUO}}
\put(65,12.5){\framebox(20,12.5){Cyc}}
\put(85,12.5){\framebox(35,12.5){SUMO}}
\put(120,12.5){\framebox(45,12.5){WordNet}}
\put(165,12.5){\framebox(45,12.5){SENSUS}}
\put(210,12.5){\framebox(30,12.5){Holes}}
\put(240,12.5){\framebox(25,12.5){Gene}}
\put(265,12.5){\framebox(35,12.5){Botany}}
\put(300,12.5){\framebox(50,12.5){Ontolingua}}
\put(40,0){\framebox(46,12.5){Enterprise}}
\put(86,0){\framebox(56,12.5){e-commerce}}
\put(142,0){\framebox(55,12.5){Government}}
\put(197,0){\framebox(50,12.5){Education}}
\put(247,0){\framebox(33,12.5){HPKB}}
\put(280,0){\framebox(70,12.5){Semantic Web}}
\put(350,0){\makebox(35,25){\Large{\ldots}}}
\end{picture}}
\end{picture}
\end{footnotesize}
\end{sffamily}
\end{center}
\caption{The IFF Architecture (detailed version)}
\label{architecture-details}
\end{figure}

\begin{table}
\begin{center}
\begin{scriptsize}
\begin{tabular}{|rl|} \hline & \\
\ssbf{Ur:}	& \underline{\ttb{IFF-UR}} [30 terms, 7 pages] \\
		& ``generic'' (category axioms) \\ 
		& \tt{metastack/UR.pdf} \\ & \\
\ssbf{Top:}	& \underline{\ttb{IFF-TCO}} [300+ terms, 60 pages] \\
		& ``very large'' (collections; finite limit axioms) \\
		& \tt{metastack/TCO.pdf} \\
		& \underline{\ttb{IFF-2CAT}} (basic 2-category theory) \\
		& \tt{metalevel/top/ontology/2-category/version20041004.pdf} \\ & \\
\ssbf{Upper:}	& \underline{\ttb{IFF-UCO}} [500+ terms, 100+ pages] \\
		& ``large'' (classes; exponents, finite limits and colimits) \\
		& \tt{metastack/UCO.pdf} \\
		& \tt{metalevel/upper/ontology/core/version20020102.pdf} \\
		& \underline{\ttb{IFF-CAT}} [220+ terms, 53 pages] \\
		& (basic category theory) \\
		& \tt{metalevel/upper/ontology/category-theory/version20020102.pdf} \\
		& \underline{\ttb{IFF-CLSN}} [280+ terms, 78 pages] \\
		& (basic information flow and formal concept analysis) \\
		& \tt{metalevel/upper/ontology/classification/version20020102.pdf} \\
		& \underline{\ttb{IFF-INS}} \\
		& (rudimentary institution theory) \\
		& \tt{work-in-progress/INS/version20031002.pdf} \\
		& \tt{work-in-progress/INS/version20041014.pdf} \\ & \\
\ssbf{Lower:}	& \underline{\ttb{IFF-LCO}} [many, many terms, 150+ pages] \\
		& ``the small'' (sets; quartets; arbitrary limits and colimits) \\
		& \tt{metalevel/lower/ontology/core/version20020515.pdf} \\
		& \tt{metalevel/lower/namespace/set/version20030402.pdf} \\
		& \underline{\ttb{IFF-CL}} [~100 terms, 36 pages] \\
		& (common logic) \\
		& \tt{metalevel/lower/namespace/scl/version20040505-obj.pdf} \\
		& \underline{\ttb{IFF-ONT}} [600+ terms, 194 pages, Language, Theory, Model, Logic namespaces] \\
		& (logic and ontology) (nontraditional FOL) (old version) \\
		& \tt{metalevel/lower/ontology/ontology/version20021205.htm} \\
		& \underline{\ttb{IFF-OO}} \\
		& (logic and ontology) (nontraditional FOL) (new version) \\
		& \tt{work-in-progress/\#IFF-OO} \\
		& \underline{\ttb{IFF-FOL}} [many, many terms, 150+ pages] \\
		& (logic and ontology) (traditional FOL) \\
		& \tt{metalevel/lower/ontology/fol/version20040101.html} \\ & \\ \hline
\end{tabular}
\end{scriptsize}
\end{center}
\caption{IFF (meta) Ontologies}
\label{meta-ontologies}
\end{table}

The IFF terminology is managed in terms of namespace prefixes
--- each namespace is given a unique prefix (with perhaps a few synonyms) in order to avoid clash of terminology\footnote{For example, the term `\ttf{morphism}' is introduced in the contexts of models and theories for first order logic to represent two distinct, but analogous, concepts. When these concepts need to be used in other contexts, the namespace prefixes `\ttf{fol.mod.mor}' and `\ttf{fol.th.mor}' could be used to distinguish them, thus resulting in the distinct representations `\ttf{fol.mod.mor:morphism}' and `\ttf{fol.th.mor:morphism}'.}.
The architecture of the IFF namespace mechanism is flat
--- namespace prefixes are like tags:
by using namespace prefixes the complete IFF terminology is the disjoint union of the terminology in the individual IFF namespaces.
The IFF architecture can be thought of as a two dimensional structure (Figure~\ref{architecture-details}) consisting of metalevels,
which are partitioned into top-level\footnote{This does not refer to the vertical dimension of the metalevel structure in Figure~\ref{architecture-details}, but instead to an implicit third dimension of the architecture\label{third:dim}.} namespaces representing basic concepts such as category (`\tts{cat}'), graph (`\tts{gph}'), or institution (`\tts{ins}').
The various levels are indexed by the natural numbers `\tts{0}', `\tts{1}', `\tts{2}', `\tts{3}', `\tts{4}', `\tts{5}', \dots, or for the first five levels by their natural language correlates `\tts{obj}', `\tts{sml}', `\tts{lrg}', `\tts{vlrg}' and `\tts{ur}', starting with the object level indexed by `\tts{0}'.

Overall, various namespaces may have the same name, since they represent the same basic concept at different metalevels.
For each basic concept the namespace axiomatization at a particular metalevel is in two parts,
one generic and the other specific:
the generic part is the specialization of the axiomatization just above it in the hierarchy;
these axiomatizations rely heavily upon the subcollection, restriction and abridgment relations for sets, functions and binary relations, respectively (Figure~\ref{subset-restriction-abridgment});
the specific part is an extension of the basic concept; it is a strictly new axiomatization. 
To locate any namespace one can use its level-concept pair.
For example, the namespace axiomatizing very large categories would be denoted by `\tts{vlrg.cat}' or `\tts{3.cat}'.
Sub-namespaces (footnote~\ref{third:dim}) (not at top-level) will need further qualification. 
For example, the sub-namespace of small graph morphisms would be denoted by `\tts{sml.gph.mor}'. 
It is assumed that each basic concept has a particular metalevel that is in common use.
These are underlined for the particular concepts in Figure~\ref{architecture-details}.
For such common use namespaces, the level notation need not be used. 
For example, the namespace that axiomatizes large categories would be denoted by `\tts{lrg.cat}' or `\tts{2.cat}', 
but more simply by `\tts{cat}'. 

The namespace mechanism has been made backward compatible, 
by allowing special namespace prefix notation that is equivalent to the general format just described.
For example, the namespace axiomatizing large categories would be denoted by the general prefixes `\tts{cat}', `\tts{lrg.cat}' or `\tts{2.cat}' or by the special prefix `\tts{CAT}' that was used in earlier versions of the \tts{IFF-CAT} axiomatization.
Hence, for any basic concept, both the general prefix and the common level need to be declared, 
and for any namespace axiomatization the special prefixes also need to be declared. 
The prefixes denoting sub-namespaces should be compatible with the top-level prefixes; 
for example, in the upper metalevel, the general prefix for the namespace of classes would be `\tts{lrg.set}' or `\tts{2.set}' with the special prefix `\tts{SET}',
whereas the general prefix for the sub-namespace of class pullbacks would be `\tts{lrg.set.lim.pbk}' with the special prefix `\tts{SET.LIM.PBK}'.
As these examples illustrate, only the lower case is needed for the general namespace prefix notation.

\section{The IFF Metastack}\label{sec-metastack}

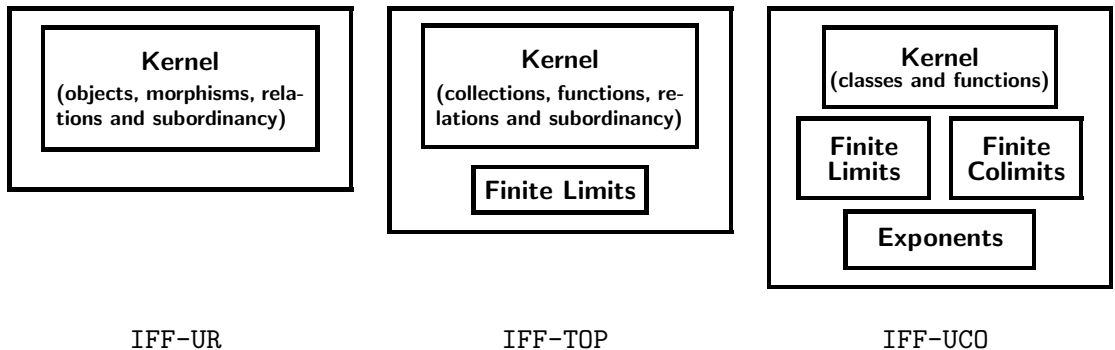
\begin{figure}
\begin{center}
\begin{sffamily}
\begin{bfseries}
\begin{footnotesize}
\begin{tabular}{ccc}
\setlength{\unitlength}{0.8pt}
\begin{picture}(160,130)(0,0)
\linethickness{1.3pt}
\put(0,46){\begin{picture}(160,84)(0,0)
\put(0,0){\framebox(160,84){}}
\put(16,19){\framebox(128,56){}}
\put(26,42){\makebox(108,36){Kernel}}
\put(21,23){\makebox(118,30){\begin{minipage}[t]{1.3in}\scriptsize{(objects, morphisms, relations and subordinancy)}\end{minipage}}}
\end{picture}}
\end{picture}
&
\setlength{\unitlength}{0.8pt}
\begin{picture}(160,130)(0,0)
\linethickness{1.3pt}
\put(0,26){\begin{picture}(160,104)(0,0)
\put(0,0){\framebox(160,104){}}
\put(16,40){\framebox(128,56){}}
\put(26,63){\makebox(108,36){Kernel}}
\put(21,43){\makebox(118,30){\begin{minipage}[t]{1.3in}\scriptsize{(collections, functions, relations and subordinancy)}\end{minipage}}}
\put(40,9){\framebox(80,20){}}
\put(50,9){\makebox(60,20){Finite Limits}}
\end{picture}}
\end{picture}
& 
\setlength{\unitlength}{0.8pt}
\begin{picture}(160,130)(0,0)
\linethickness{1.3pt}
\put(0,0){\framebox(160,130){}}
\put(26,85){\framebox(108,36){}}
\put(26,90){\makebox(108,36){Kernel}}
\put(26,78){\makebox(108,36){\scriptsize{(classes and functions)}}}
\put(14,42){\framebox(60,36){}}
\put(14,48){\makebox(60,36){Finite}}
\put(14,36){\makebox(60,36){Limits}}
\put(86,42){\framebox(60,36){}}
\put(86,48){\makebox(60,36){Finite}}
\put(86,36){\makebox(60,36){Colimits}}
\put(36.25,9){\framebox(87.5,25){}}
\put(36.25,9){\makebox(87.5,25){Exponents}}
\end{picture}
\\ \\
\bfseries{\small{\ttfamily{IFF-UR}}}
& \bfseries{\small{\ttfamily{IFF-TOP}}} 
& \bfseries{\small{\ttfamily{IFF-UCO}}}
\end{tabular}
\end{footnotesize}
\end{bfseries}
\end{sffamily}
\end{center}
\caption{Core Architecture (iconic version)}
\label{core-architecture-iconic}
\end{figure}

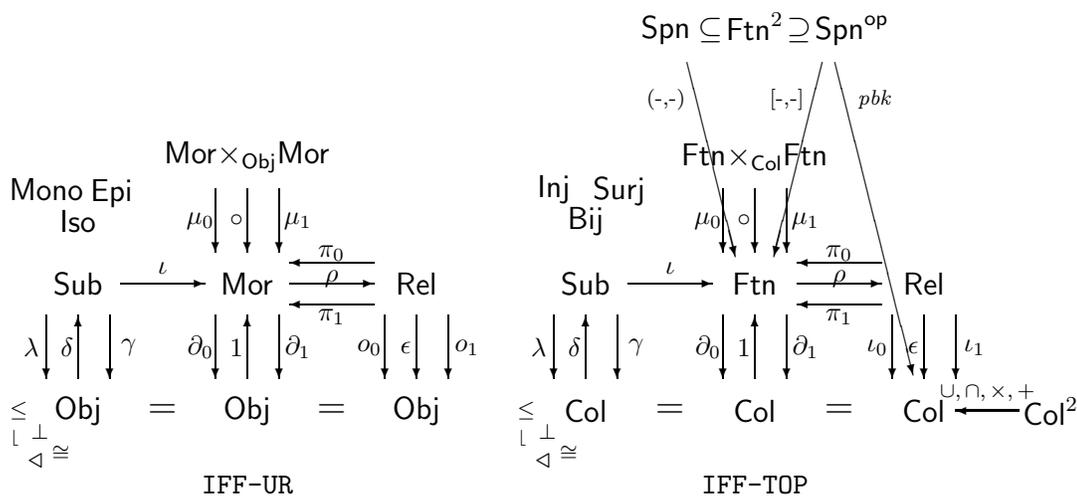
\begin{figure}
\begin{center}
\setlength{\unitlength}{0.8pt}
\begin{picture}(400,225)(0,0)
\put(0,0){\begin{picture}(160,150)(0,-30)
\put(40,-65){\makebox(80,60){\bfseries\small{\ttfamily{IFF-UR}}}}
\put(-35,90){\makebox(40,30){$\mathsf{Mono}$}}
\put(-4,88){\makebox(40,30){$\mathsf{Epi}$}}
\put(-20,75){\makebox(40,30){$\mathsf{Iso}$}}
\put(-20,45){\makebox(40,30){$\mathsf{Sub}$}}
\put(-20,-15){\makebox(40,30){$\mathsf{Obj}$}}
\put(60,45){\makebox(40,30){$\mathsf{Mor}$}}
\put(60,-15){\makebox(40,30){$\mathsf{Obj}$}}
\put(40,105){\makebox(80,30){$\mathsf{Mor} {\times}_{\mathsf{Obj}} \mathsf{Mor}$}}
\put(140,45){\makebox(40,30){$\mathsf{Rel}$}}
\put(140,-15){\makebox(40,30){$\mathsf{Obj}$}}
\put(110,-7.5){\makebox(20,15){\large{$=$}}}
\put(-48,-15){\makebox(40,30){\scriptsize{$\leq$}}}
\put(-38,-25){\makebox(40,30){\scriptsize{$\perp$}}}
\put(-28,-35){\makebox(40,30){\scriptsize{$\cong$}}}
\put(-50,-28){\makebox(40,30){\tiny{$\lfloor$}}}
\put(-40,-39){\makebox(40,30){\small{$\triangleleft$}}}
\put(-32,22.5){\makebox(20,15){\footnotesize{$\lambda$}}}
\put(-15,22.5){\makebox(20,15){\footnotesize{$\delta$}}}
\put(14,22.5){\makebox(20,15){\footnotesize{$\gamma$}}}
\put(-15,45){\vector(0,-1){30}}
\put(0,15){\vector(0,1){30}}
\put(15,45){\vector(0,-1){30}}
\put(20,60){\vector(1,0){40}}
\put(30,60){\makebox(20,15){\footnotesize{$\iota$}}}
\put(30,-7.5){\makebox(20,15){\large{$=$}}}
\put(140,70){\vector(-1,0){40}}
\put(100,60){\vector(1,0){40}}
\put(140,50){\vector(-1,0){40}}
\put(110,67){\makebox(20,15){\footnotesize{$\pi_{0}$}}}
\put(110,55){\makebox(20,15){\footnotesize{$\rho$}}}
\put(110,37){\makebox(20,15){\footnotesize{$\pi_{1}$}}}
\put(48,22.5){\makebox(20,15){\footnotesize{$\partial_{0}$}}}
\put(65,22.5){\makebox(20,15){\footnotesize{$1$}}}
\put(94,22.5){\makebox(20,15){\footnotesize{$\partial_{1}$}}}
\put(65,45){\vector(0,-1){30}}
\put(80,15){\vector(0,1){30}}
\put(95,45){\vector(0,-1){30}}
\put(48,82.5){\makebox(20,15){\footnotesize{$\mu_{0}$}}}
\put(65,82.5){\makebox(20,15){\footnotesize{$\circ$}}}
\put(94,82.5){\makebox(20,15){\footnotesize{$\mu_{1}$}}}
\put(65,105){\vector(0,-1){30}}
\put(80,105){\vector(0,-1){30}}
\put(95,105){\vector(0,-1){30}}
\put(128,22.5){\makebox(20,15){\footnotesize{$o_{0}$}}}
\put(145,22.5){\makebox(20,15){\footnotesize{$\epsilon$}}}
\put(174,22.5){\makebox(20,15){\footnotesize{$o_{1}$}}}
\put(145,45){\vector(0,-1){30}}
\put(160,45){\vector(0,-1){30}}
\put(175,45){\vector(0,-1){30}}
\end{picture}}
\put(240,0){\begin{picture}(160,150)(0,-30)
\put(40,-65){\makebox(80,60){\bfseries\small{\ttfamily{IFF-TOP}}}}
\put(-2,165){\makebox(80,30){$\mathsf{Spn}$}}
\put(19,165){\makebox(80,30){$\subseteq$}}
\put(40,167){\makebox(80,30){$\mathsf{Ftn}^{2}$}}
\put(61,165){\makebox(80,30){$\supseteq$}}
\put(86,165){\makebox(80,30){$\mathsf{Spn}^{{\mathsf{op}}}$}}
\put(28,140){\makebox(20,15){\scriptsize{(-,-)}}}
\put(85,140){\makebox(20,15){\scriptsize{[-,-]}}}
\put(127,140){\makebox(20,15){\scriptsize{\emph{pbk}}}}
\put(48,165){\vector(1,-4){23}}
\put(112,165){\vector(-1,-4){23}}
\put(118,165){\vector(1,-4){37}}
\put(-35,90){\makebox(40,30){$\mathsf{Inj}$}}
\put(-4,89){\makebox(40,30){$\mathsf{Surj}$}}
\put(-20,75){\makebox(40,30){$\mathsf{Bij}$}}
\put(-20,45){\makebox(40,30){$\mathsf{Sub}$}}
\put(-20,-15){\makebox(40,30){$\mathsf{Col}$}}
\put(60,45){\makebox(40,30){$\mathsf{Ftn}$}}
\put(60,-15){\makebox(40,30){$\mathsf{Col}$}}
\put(40,105){\makebox(80,30){$\mathsf{Ftn} {\times}_{\mathsf{Col}} \mathsf{Ftn}$}}
\put(140,45){\makebox(40,30){$\mathsf{Rel}$}}
\put(140,-15){\makebox(40,30){$\mathsf{Col}$}}
\put(200,-15){\makebox(40,30){$\mathsf{Col}^{2}$}}
\put(110,-7.5){\makebox(20,15){\large{$=$}}}
\put(-48,-15){\makebox(40,30){\scriptsize{$\leq$}}}
\put(-38,-25){\makebox(40,30){\scriptsize{$\perp$}}}
\put(-28,-35){\makebox(40,30){\scriptsize{$\cong$}}}
\put(-50,-28){\makebox(40,30){\tiny{$\lfloor$}}}
\put(-40,-39){\makebox(40,30){\small{$\triangleleft$}}}
\put(205,0){\vector(-1,0){30}}
\put(180,1){\makebox(20,15){\scriptsize{$\cup$,\,$\cap$,\,$\times$,\,$+$}}}
\put(-32,22.5){\makebox(20,15){\footnotesize{$\lambda$}}}
\put(-15,22.5){\makebox(20,15){\footnotesize{$\delta$}}}
\put(14,22.5){\makebox(20,15){\footnotesize{$\gamma$}}}
\put(-15,45){\vector(0,-1){30}}
\put(0,15){\vector(0,1){30}}
\put(15,45){\vector(0,-1){30}}
\put(20,60){\vector(1,0){40}}
\put(30,60){\makebox(20,15){\footnotesize{$\iota$}}}
\put(30,-7.5){\makebox(20,15){\large{$=$}}}
\put(140,70){\vector(-1,0){40}}
\put(100,60){\vector(1,0){40}}
\put(140,50){\vector(-1,0){40}}
\put(110,67){\makebox(20,15){\footnotesize{$\pi_{0}$}}}
\put(110,55){\makebox(20,15){\footnotesize{$\rho$}}}
\put(110,37){\makebox(20,15){\footnotesize{$\pi_{1}$}}}
\put(48,22.5){\makebox(20,15){\footnotesize{$\partial_{0}$}}}
\put(65,22.5){\makebox(20,15){\footnotesize{$1$}}}
\put(94,22.5){\makebox(20,15){\footnotesize{$\partial_{1}$}}}
\put(65,45){\vector(0,-1){30}}
\put(80,15){\vector(0,1){30}}
\put(95,45){\vector(0,-1){30}}
\put(48,82.5){\makebox(20,15){\footnotesize{$\mu_{0}$}}}
\put(65,82.5){\makebox(20,15){\footnotesize{$\circ$}}}
\put(94,82.5){\makebox(20,15){\footnotesize{$\mu_{1}$}}}
\put(65,105){\vector(0,-1){30}}
\put(80,105){\vector(0,-1){30}}
\put(95,105){\vector(0,-1){30}}
\put(128,22.5){\makebox(20,15){\footnotesize{$\iota_{0}$}}}
\put(145,22.5){\makebox(20,15){\footnotesize{$\epsilon$}}}
\put(174,22.5){\makebox(20,15){\footnotesize{$\iota_{1}$}}}
\put(145,45){\vector(0,-1){30}}
\put(160,45){\vector(0,-1){30}}
\put(175,45){\vector(0,-1){30}}
\end{picture}}
\end{picture}
\end{center}
\caption{Core Architecture (detailed version)}
\label{core-architecture-details}
\end{figure}

\begin{table}
\begin{center}
\begin{scriptsize}
\begin{tabular}{|r@{\hspace{3pt}}c@{\hspace{3pt}}l||r@{\hspace{3pt}}c@{\hspace{3pt}}l||r@{\hspace{3pt}}c@{\hspace{3pt}}l|}
\multicolumn{3}{l}{\sss{Set}} 
& \multicolumn{3}{l}{\sss{Function}} 
& \multicolumn{3}{l}{\sss{Relation}} \\ \hline
&&`\tt{thing}'&&&&&&\\ \hline\hline
$\mathsf{Obj}$&$\doteq$&`\tt{object}'&&&&$\leq$&$\doteq$&`\tt{subobject}'\\
&&&&&&$\perp$&$\doteq$&`\tt{disjoint}'\\
&&&&&&$\cong$&$\doteq$&`\tt{isomorphic}'\\
&&&&&&$\lfloor$&$\doteq$&`\tt{restriction}'\\
&&&&&&\normalsize{$\triangleleft$}&$\doteq$&`\tt{abridgment}'\\ \hline\hline
$\mathsf{Mor}$&$\doteq$&`\tt{morphism}'&$\partial_{0}$, $\partial_{1}$&$\doteq$&`\tt{source}', `\tt{target}'&&&`\tt{restriction}'\\
&&&$\rho$&$\doteq$&`\tt{mor2rel}'&$\mathsf{Mono}$&$\doteq$&`\tt{monomorphism}'\\
$\mathsf{Mor} {\times}_{\mathsf{Obj}} \mathsf{Mor}$&$\doteq$&`\tt{morphism}
&$\mu_{0}$, $\mu_{1}$&$\doteq$&`\tt{morphism0}', `\tt{morphism1}'&$\mathsf{Epi}$&$\doteq$&`\tt{epimorphism}'\\
&&\tt{-morphism}'&$\circ$, $1$&$\doteq$&`\tt{composition}', `\tt{identity}'&$\mathsf{Iso}$&$\doteq$&`\tt{isomorphism}'\\ \hline\hline
$\mathsf{Rel}$&$\doteq$&`\tt{relation}'&$o_{0}$, $o_{1}$&$\doteq$&`\tt{object0}', `\tt{object1}'&&&`\tt{abridgment}'\\
&&&$\epsilon$&$\doteq$&`\tt{extent}'&&&\\
&&&$\pi_{0}$, $\pi_{1}$&$\doteq$&`\tt{projection0}', `\tt{projection1}'&&&\\ \hline\hline
$\mathsf{Sub}$&$\doteq$&`\tt{subordinate}'&$\lambda$, $\gamma$&$\doteq$&`\tt{lesser}', `\tt{greater}'&&&\\
&&&$\iota$, $\delta$&$\doteq$&`\tt{inclusion}', `\tt{reflex}'&&&\\ \hline
\end{tabular}
\end{scriptsize}
\end{center}
\caption{\tts{IFF-UR} correspondences}
\label{fig-ur-terms}
\end{table}

The left sides of Figures~\ref{core-architecture-iconic} and~\ref{core-architecture-details}
illustrate the architecture for the IFF Ur (meta) Ontology \tts{IFF-UR},
a tiny ontology at the very top of the metastack, the IFF core metalevel hierarchy.
Its purpose is to provide an interface between the IFF logical metashell and other metalevel axiomatizations.
Principally, it does this by servicing the top metalevel.
A design constraint is that ``things are opaque'';
that is, the detailed state of affairs ``inside'' anything is unknown.
Such detail is to be provided by a using ontology,
such as the IFF Top Core (meta) Ontology (\tts{IFF-TCO}),
the IFF 2-Category (meta) Ontology (\tts{IFF-2CAT}),
and the future IFF Double-Category (meta) Ontology (\tts{IFF-DBLCAT}).
The \tts{IFF-UR} contribution to the \tts{meta-ur} metalanguage is a simple framework
--- the \tts{IFF-TCO}, \tts{IFF-2CAT} and \tts{IFF-DCAT} fill it out.
Table~\ref{fig-ur-terms} shows various correspondences between the elements on the left side of Figure~\ref{core-architecture-details} and the terminology of \tts{IFF-UR}
(the \tts{IFF-UR}-term `\tts{thing}' is not represented in Figure~\ref{core-architecture-details}).
The \tts{IFF-UR} terminology consists of only 30 terms: 6 generic sets; 16 generic functions;
and 8 generic relations, consisting of 3 binary endorelation on objects, 1 binary and 3 unary relations on morphisms, and 1 binary relation on relations.
The subobject, restriction and abridgment binary relations are important for the preservation properties of the metastack.

The middle of Figure~\ref{core-architecture-iconic} and the right side of Figure~\ref{core-architecture-details} 
illustrate the architecture for the IFF Top Core (meta) Ontology \tts{IFF-TCO},
which is situated in the top metalevel
--- the highest level of the IFF metastack other than Ur.
It contains rudimentary (fundamental) namespaces for collections, functions, relations and finite limits.
The \tts{IFF-TCO} provides an adequate foundation for representing ontologies in general and for defining other metalevel ontologies in particular.
The fact that the \tts{IFF-TCO} includes the specialization of the \tts{IFF-UR} is evident from the isomorphic embedding of the left side into the right side of Figure~\ref{core-architecture-details}. 
The \tts{IFF-UR} follows Mac Lane's beginning axiomatization for category theory \cite{maclane:71} in that it introduces terminology and provides an axiomatization for this terminology,
but it does not give a formal interpretation using set theory
--- it only gives an informal, intuitive interpretation.
The \tts{IFF-TCO} initiates such a formal interpretation.
The IFF Upper and Lower Core (meta) Ontologies, \tts{IFF-UCO} and \tts{IFF-LCO}, complete such a formal interpretation.

The \tts{IFF-TCO} axiomatization contains 317 terms (276 concepts or non-identical terms),
partitioned according to whether the term is a basic term (collections, relations or functions),
a diagram term or a limit term.
Ignoring the 10 terms used to designate the four indexing collections,
there are a total of 307 terms partitioned into 58 basic terms designating 55 basic concepts dealing with collections, (partial) functions and relations,
105 terms designating 97 concepts of finite diagrams,
and 144 terms designating 120 concepts of finite limits proper.
Although the \tts{IFF-TCO} axiomatizates finite limits,
the current version does not represent finiteness,
but instead explicitly represents the several finite shapes needed.
A future version of the \tts{IFF-TCO} may experiment with Dedekind's abstract definition of finiteness \cite{lawvere:rosebrugh:03}.
Terminology has been placed in the IFF Top Core Ontology only when it is needed in the IFF upper metalevel.
All upper metalevel ontologies import and use, either directly or indirectly, the IFF Top Core Ontology.
This includes the upper core, upper graph and upper category theory meta-ontologies.

In the third phase,
a revised version of the IFF Upper Core (meta) Ontology (\tts{IFF-UCO}) will be designed.
The right side of Figure~\ref{core-architecture-iconic}
illustrates the architecture for the \tts{IFF-TCO}.
This axiomatization will be in adjunctive form.
The \tts{IFF-UCO} plays a central role in the IFF axiomatization
--- it is the most referenced ontology at the metalevel.
The \tts{IFF-UCO} kernel consists of the axiomatization for classes and functions.
At the upper metalevel the axiomatization for relations, which includes order-theoretic concepts,
has branched off into an ontology in its own right.
The kernel and finite limits namespace axiomatizations are the specializations of those in the \tts{IFF-TCO},
and as such rely heavily upon the subcollection, restriction and abridgment relations for collections, functions and binary relations, respectively (Figure~\ref{subset-restriction-abridgment}).
The finite colimits namespace is a categorical dual to the finite limits namespace.
The exponents namespace axiomatization is new.

\section{The IFF Lower Metalevel}\label{sec-lower-metalevel}

\begin{figure}
\begin{center}
\begin{sffamily}
\begin{bfseries}
\begin{picture}(126,130)(0,0)
\linethickness{1.3pt}
\put(0,0){\begin{picture}(126,126)(0,0)
\put(0,0){\framebox(126,126){}}
\put(4,98){\makebox(118,36)[l]{\tt{trm.lang}}}
\put(4,83){\makebox(118,36)[l]{category namespace}}
\put(11,52){\framebox(106,36){}}
\put(14,58){\makebox(98,36)[l]{\ttf{trm.lang.obj}}}
\put(14,45){\makebox(98,36)[l]{\footnotesize{object namespace}}}
\put(11,10){\framebox(106,36){}}
\put(14,16){\makebox(98,36)[l]{\ttf{trm.lang.mor}}}
\put(14,3){\makebox(98,36)[l]{\footnotesize{morphism namespace}}}
\end{picture}}
\end{picture}
\end{bfseries}
\end{sffamily}
\end{center}
\caption{Lower Core Format}
\label{lower-core-format}
\end{figure}
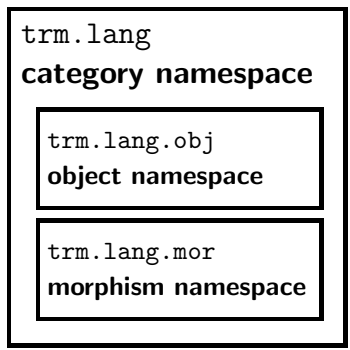
\begin{figure}
\begin{center}
\setlength{\unitlength}{0.8pt}
\begin{picture}(330,175)(0,-20)
\put(0,80){\begin{picture}(120,40)(0,0)
\put(-16,31){\dashbox{3}(164,25){}}
\put(-25,55){\makebox(60,20)[l]{{\textsc{Term}}}}
\put(-15,30){\makebox(30,20){$\mathsf{case}$}}
\put(45,30){\makebox(30,20){$\mathsf{trm}$}}
\put(105,30){\makebox(30,20){$\mathsf{ftn}\otimes\mathsf{tpl}$}}
\put(45,-10){\makebox(30,20){$\mathsf{trm}$}}
\put(105,-10){\makebox(30,20){$\mathsf{trm}\otimes\mathsf{tpl}$}}
\put(14,40){\makebox(30,20){\scriptsize{$\mathsf{elem}$}}}
\put(15,30){\makebox(30,20){\large{$\Rightarrow$}}}
\put(68,40){\makebox(30,20){\scriptsize{$\mathsf{subst}$}}}
\put(67,30){\makebox(30,20){\large{$\Leftarrow$}}}
\put(68,-1){\makebox(30,20){\scriptsize{$\mathsf{subst}$}}}
\put(66,-11){\makebox(30,20){\large{$\Leftarrow$}}}
\put(45,10){\makebox(30,20){\large{$=$}}}
\put(128,12){\makebox(30,20){\scriptsize{$\epsilon \otimes 1_{\mathsf{tpl}}$}}}
\put(105,10){\makebox(30,20){\large{$\Downarrow$}}}
\end{picture}}
\put(0,0){\begin{picture}(120,40)(0,0)
\put(-18,-16){\dashbox{3}(156,25){}}
\put(-25,55){\makebox(60,20)[l]{{\textsc{Tuple}}}}
\put(-55,10){\makebox(30,20){$\mathsf{case}$}}
\put(-26,11){\makebox(30,20){\footnotesize{$\supseteq$}}}
\put(145,10){\makebox(30,20){$\mathsf{ftn}$}}
\put(-15,30){\makebox(30,20){$\mathsf{var}$}}
\put(-15,-10){\makebox(30,20){$\mathsf{var}{\circ}{\wp}$}}
\put(45,30){\makebox(30,20){$\mathsf{var} \times \mathsf{trm}$}}
\put(105,30){\makebox(30,20){$\mathsf{trm}$}}
\put(45,-10){\makebox(30,20){$\mathsf{tpl}$}}
\put(105,-10){\makebox(30,20){$\mathsf{var}{\circ}{\wp}$}}
\put(7,39){\makebox(30,20){\scriptsize{$\pi_{0}$}}}
\put(5,30){\makebox(30,20){\large{$\Leftarrow$}}}
\put(84,39){\makebox(30,20){\scriptsize{$\pi_{1}$}}}
\put(84,30){\makebox(30,20){\large{$\Rightarrow$}}}
\put(-43,31){\makebox(30,20){\scriptsize{$\mathsf{proj}$}}}
\put(-38,20){\makebox(30,20){\large{$\Rightarrow$}}}
\put(-38,0){\makebox(30,20){\large{$\Rightarrow$}}}
\put(-45,-7){\makebox(30,20){\scriptsize{$\mathsf{indic}$}}}
\put(131,30){\makebox(30,20){\scriptsize{$\epsilon$}}}
\put(128,20){\makebox(30,20){\large{$\Leftarrow$}}}
\put(128,0){\makebox(30,20){\large{$\Leftarrow$}}}
\put(131,-7){\makebox(30,20){\tiny{$\#$}}}
\put(20,-19){\makebox(30,20){\scriptsize{$\S$}}}
\put(18,-10){\makebox(30,20){\large{$\Leftarrow$}}}
\put(70,-19){\makebox(30,20){\tiny{$\#$}}}
\put(70,-10){\makebox(30,20){\large{$\Rightarrow$}}}
\put(-3,10){\makebox(30,20){\scriptsize{\{-\}}}}
\put(-15,10){\makebox(30,20){\large{$\Downarrow$}}}
\put(33,10){\makebox(30,20){\scriptsize{\{-\}}}}
\put(45,10){\makebox(30,20){\large{$\Downarrow$}}}
\put(95,10){\makebox(30,20){\tiny{$\#$}}}
\put(105,10){\makebox(30,20){\large{$\Downarrow$}}}
\end{picture}}
\put(210,0){\begin{picture}(120,120)(0,0)
\put(-23,-10){\dashbox{3}(165,145){}}
\multiput(-67,-4)(6,0){8}{\line(-1,0){3}}
\put(-25,135){\makebox(60,20)[l]{{\textsc{Lawvere}}}}
\put(-15,110){\makebox(30,20){$\mathsf{var}{\circ}{\wp}$}}
\put(45,110){\makebox(30,20){\large{$=$}}}
\put(105,110){\makebox(30,20){$\mathsf{var}{\circ}{\wp}$}}
\put(-15,70){\makebox(30,20){$\mathsf{tpl}$}}
\put(45,70){\makebox(30,20){$\mathsf{tpl} \otimes \mathsf{tpl}$}}
\put(105,70){\makebox(30,20){$\mathsf{tpl}$}}
\put(-15,30){\makebox(30,20){$\mathsf{var}{\circ}{\wp}$}}
\put(45,30){\makebox(30,20){$\mathsf{tpl}$}}
\put(105,30){\makebox(30,20){$\mathsf{var}{\circ}{\wp}$}}
\put(45,-10){\makebox(30,20){$\mathsf{var}{\circ}{\wp}$}}
\put(-25,92){\makebox(30,20){\tiny{$\#$}}}
\put(-15,92){\makebox(30,20){\large{$\Uparrow$}}}
\put(115,92){\makebox(30,20){\scriptsize{$\S$}}}
\put(105,92){\makebox(30,20){\large{$\Uparrow$}}}
\put(8,81){\makebox(30,20){\scriptsize{$0^{\mathsf{th}}$}}}
\put(6,70){\makebox(30,20){\large{$\Leftarrow$}}}
\put(82,81){\makebox(30,20){\scriptsize{$1^{\mathsf{st}}$}}}
\put(82,70){\makebox(30,20){\large{$\Rightarrow$}}}
\put(20,21){\makebox(30,20){\scriptsize{$\S$}}}
\put(18,30){\makebox(30,20){\large{$\Leftarrow$}}}
\put(70,21){\makebox(30,20){\tiny{$\#$}}}
\put(70,30){\makebox(30,20){\large{$\Rightarrow$}}}
\put(-25,50){\makebox(30,20){\scriptsize{$\S$}}}
\put(-15,50){\makebox(30,20){\large{$\Downarrow$}}}
\put(34,51){\makebox(30,20){\scriptsize{$\circ$}}}
\put(45,50){\makebox(30,20){\large{$\Downarrow$}}}
\put(115,50){\makebox(30,20){\tiny{$\#$}}}
\put(105,50){\makebox(30,20){\large{$\Downarrow$}}}
\put(5,10){\makebox(30,20){\large{$=$}}}
\put(35,8){\makebox(30,20){\scriptsize{$1$}}}
\put(45,10){\makebox(30,20){\large{$\Uparrow$}}}
\put(85,10){\makebox(30,20){\large{$=$}}}
\end{picture}}
\put(175,-45){\begin{minipage}[t]{165pt}
\begin{scriptsize}
\begin{tabular}{r@{\hspace{5pt}}c@{\hspace{5pt}}l}
$\epsilon$  &=& function as term embedding \\
\tiny{$\#$} &=& function/term/tuple arity  \\
$\S$        &=& tuple index                \\  
\end{tabular}
\end{scriptsize}
\end{minipage}}
\end{picture}
\end{center}
\caption{Term Language Architecture: \newline {\sffamily Set}-valued functors and natural transformations \newline on the category {\sffamily Trm-Lang}}
\label{term-architecture}
\end{figure}
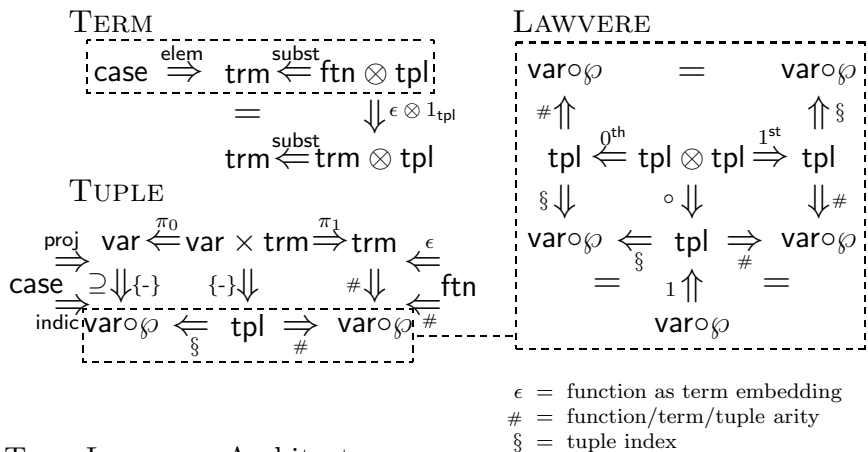

Each lower-level meta-ontology in the IFF is concentrated around a single category (with its associated functors, natural transformations and monads),
and hence the format of each lower-level meta-ontology has a specific category-theoretic format consisting of an ``outer'' category namespace and two ``inner'' namespaces for objects and morphisms.
In Figure~\ref{lower-core-format} we illustrate this format with the category {\sffamily Trm-Lang} of term languages and term language morphisms\footnote{A term language is simple 
--- it consists of
a set of (indexing) variables,
a set of function types or symbols, and
an arity function from function symbols to indicia (subsets of variables).
A term language morphism is also simple
--- it consists of
a variable bijection and a function between function symbol sets,
which form a set quartet between source/target arity functions (preserves arity).
We could have ignored variables, indicia and arity for languages, and just used natural numbers and valence instead, with language morphisms required to preserve valence.
In fact, the classic Lawvere category does this
--- it uses natural numbers as objects, in place of indicia, and uses term sequences in place of term tuples. However, we used variables for at least four reasons.
(1) Standards such as the common logic standard (CL) use a set of variables as part of their lexicons.
(2) We did not want to rely upon sequences and a natural numbers axiomatization.
(3) The use of variable-type pairs for arguments in function declaration is common in programming languages.
(4) We want the expression operator, which requires variables and arities as part of the formulation of expressions, to be a monad. It is very easy to add natural numbers, valence and term sequences, thus presenting traditional first order logic.},
which is axiomatized in the term language namespace \tts{trm.lang} of the IFF Term (meta) Ontology \tts{IFF-TRM}.
\begin{table}
\begin{center}
\begin{scriptsize}
\begin{tabular}{|p{4.1in}|} \hline
\\ 
\ttbf{trm.lang}: ``\emph{There is a (large) category} {\sffamily Trm-Lang} \emph{of term languages. The object class of} {\sffamily Trm-Lang} \emph{is the class of all term languages. The morphism class of} {\sffamily Trm-Lang} \emph{is the class of all term language morphisms. Composition in} {\sffamily Trm-Lang} \emph{is composition of term language morphisms.}''
\begin{verbatim}
(cat:category language)
(= (cat:object language) trm.lang.obj:object)
(= (cat:morphism language) trm.lang.mor:morphism)
(= (cat:source language) trm.lang.mor:source)
(= (cat:target language) trm.lang.mor:target)
(= (cat:composable language) trm.lang.mor:composable)
(= (cat:composition language) trm.lang.mor:composition)
(= (cat:identity language) trm.lang.mor:identity)
\end{verbatim} \\ \hline\hline
\\
\ttbf{trm.lang}: ``\emph{There is a function symbol functor from the (large) category of term languages to the (large) category of sets}
$\mathsf{ftn} : \mathsf{Trm}\mbox{-}\mathsf{Lang} \rightarrow \mathsf{Set}$.
\emph{There is an arity natural transformation}
{\tiny $\#$}$\,: \mathsf{ftn} \Rightarrow \mathsf{var} \circ \wp : \mathsf{Trm}\mbox{-}\mathsf{Lang} \rightarrow \mathsf{Set}$,
\emph{whose} $L^{\mathrm{th}}$ \emph{component for any term language $L$ is the arity function for $L$} .''
\begin{verbatim}
(func:functor function)
(= (func:source function) language)
(= (func:target function) set:set)
(= (func:object function) trm.lang.obj:function)
(= (func:morphism function) trm.lang.mor:function)

(nat:natural-transformation function-arity)
(= (nat:source-category function-arity) language)
(= (nat:target-category function-arity) set:set)
(= (nat:source-functor function-arity) function)
(= (nat:target-functor function-arity) indicia)
(= (nat:component function-arity) trm.lang.obj:function-arity)
\end{verbatim}
\\ \hline\hline
\\
\ttbf{trm.lang}: ``\emph{The Lawvere construction is a functor}
$\mathsf{law} : \mathsf{Trm}\mbox{-}\mathsf{Lang} \rightarrow \mathsf{Cat}$ 
\emph{from the (large) category of term languages to the (large) category of small categories:
for any term language} $L$, $\mathsf{law}(L)$ \emph{is a small category with coproducts and for any term language morphism} $f : L_{1} \rightarrow L_{2}$,
$\mathsf{law}(f) : \mathsf{law}(L_{1}) \rightarrow \mathsf{law}(L_{2})$ \emph{is a functor between small categories that preserves coproducts.}''
\begin{verbatim}
(func:functor lawvere)
(= (func:source lawvere) language)
(= (func:target lawvere) sml.cat:category)
(= (func:object lawvere) trm.lang.obj.tpl:lawvere)
(= (func:morphism lawvere) trm.lang.mor.tpl:lawvere)
\end{verbatim}
\\ \hline
\end{tabular}
\end{scriptsize}
\end{center}
\caption{Code for {\ttfamily trm-lang}}
\label{term-code}
\end{table}
The category namespace of \tts{trm.lang} is completely compliant with the categorical-design principle.
This is illustrated in Table~\ref{term-code},
which lists the logical code for
the category {\sffamily Trm-Lang},
the functor $\mathsf{ftn} : \mathsf{Trm}\mbox{-}\mathsf{Lang} \rightarrow \mathsf{Set}$,
the arity natural transformation
{\footnotesize $\#$}$\,: \mathsf{ftn} \Rightarrow \mathsf{var} \circ \wp : \mathsf{Trm}\mbox{-}\mathsf{Lang} \rightarrow \mathsf{Set}$,
and the Lawvere construction $\mathsf{law} : \mathsf{Trm}\mbox{-}\mathsf{Lang} \rightarrow \mathsf{Cat}$\footnote{The functor $\mathsf{ftn}$ and the natural transformation {\footnotesize $\#$} are explicitly represented in Figure~\ref{term-architecture},
whereas {\sffamily Trm-Lang} is the ambient category,
and $\mathsf{law}$ is represented by the Lawvere sub-diagram.}.
The ``outer'' category namespace of \tts{trm.lang} uses terminology from the \tts{meta-upper} metalanguage,
plus the lower metalevel terminology from the object and morphism namespaces of \tts{trm.lang}.
The latter are only partially (80--90\%) compliant with the categorical-design principle.
Part of the goal for phase 3 is to bring these ``inner'' namespaces toward 100\% compliance.

The basic architecture of term languages is illustrated in Figure~\ref{term-architecture},
which consists of three sub-diagrams
--- the term sub-diagram (upper left), the term tuple sub-diagram (lower left) and the Lawvere sub-diagram (right). The term sub-diagram illustrates the fixpoint solution for terms,
the term tuple sub-diagram illustrates the embedding structures for term tuples,
and the Lawvere sub-diagram illustrates the Lawvere category $\mathsf{law}(L)$ of a term language $L$,
which has indicia as objects, term tuples as morphisms, and substitution as composition.
The Lawvere construction is a collection of (small) categories and functors indexed by term languages and term language morphisms.
Abstractly, the Lawvere construction is a (small) category object in the (large) category of functors and natural transformations between the categories {\sffamily Trm-Lang} and {\sffamily Set}.
In Figure~\ref{term-architecture}, the nodes represent basic functors from {\sffamily Trm-Lang} to {\sffamily Set},
and the edges represent natural transformations between these basic functors.
There are five simple functors {\sffamily var}, {\sffamily ftn}, {\sffamily case}, {\sffamily trm} and {\sffamily tpl}, for variables, function symbols, variable cases, terms and term tuples, respectively.
The first three are basic and the last two are inductively defined.
Based on these, there are five composite functors:
$\mathsf{var}{\circ}{\wp}$, $\mathsf{var}{\times}\mathsf{trm}$, $\mathsf{ftn}\otimes\mathsf{tpl}$,
$\mathsf{trm}\otimes\mathsf{tpl}$, $\mathsf{tpl} \otimes \mathsf{tpl}$,
for indicia (variable subsets), indexed terms (variable-term pairs), substitutable function-tuple pairs, substitutable term-tuple pairs and composable tuple-tuple pairs, respectively.
The '$\otimes$' symbol refers to a matched Cartesian product
--- the arity of the first (function, term or tuple) matches the index of the second (tuple).

\section{The IFF Metalanguages}\label{sec-metalanguages}

Table~\ref{example-terms} lists selected examples of IFF terminology organized along metalevels.
In general, the terminology introduced at any level uses the terminology in the same or higher metalevels.
Object level ontologies use lower metalevel terminology and functionality.
For example, 
the terminology and functionality introduced and axiomatized in the lower metalevel IFF theory namespace (\ttbs{trm}),
which is part of the IFF First Order Logic (meta) Ontology (\tts{IFF-FOL}),
could be used to organize and axiomatize an E-commerce ontology in the object level.
Lower metalevel namespaces and meta-ontologies use upper level terminology and functionality.
For example, 
the terminology and functionality introduced and axiomatized in the upper metalevel IFF Category Theory (meta) Ontology (\ttbs{IFF-CAT}) is used to organize and axiomatize the lower metalevel IFF language namespace (\ttbs{lang})
that is part of the \tts{IFF-FOL} meta-ontology.
Upper meta\-level namespaces and meta-ontologies use top metalevel terminology and functionality.
For example, 
the terminology and functionality introduced and axiomatized in the top metalevel IFF Top Core (meta) Ontology (\ttbs{IFF-TCO}) is used to organize and axiomatize the \ttbs{IFF-CAT} meta-ontology. 
\begin{table}
\begin{scriptsize}
\begin{tabular}{r|p{6.5cm}|p{6.5cm}|}
\multicolumn{1}{l}{}
& \multicolumn{1}{p{6.5cm}}{\mbox{\rule[-0.2cm]{0cm}{0.4cm}\ssbf{Core}}}
& \multicolumn{1}{p{6.5cm}}{\ssbf{Periphery}} \\ \cline{2-3}
\ttb{metashell}	
& \multicolumn{2}{p{13cm}|}{`\tt{and}', `\tt{or}', `\tt{implies}', `\tt{iff}', `\tt{forall}', `\tt{not}', `\tt{forall}', `\tt{exists}'}	
\\ \cline{2-3}
\ttb{meta-ur}	
& \raggedright `\tt{object}', `\tt{morphism}', `\tt{relation}', `\tt{subobject}', `\tt{restriction}', `\tt{abridgment}', `\tt{composition}', `\tt{identity}' 
& \\ \cline{2-3}
\ttb{meta-top}	
& \raggedright `\tt{collection}', `\tt{subcollection}', `\tt{function}', `\tt{composition}', `\tt{identity}', `\tt{restriction}', `\tt{relation}', `\tt{abridgment}'
& \\ \cline{2-2}	
& \raggedright `\tt{collection-pair}', `\tt{parallel-pair}', `\tt{opspan}', `\tt{binary-product}', `\tt{equalizer}', `\tt{pullback}'	
& \\ \cline{2-3}
\ttb{meta-upper}
& \raggedright `\tt{class}', `\tt{function}', `\tt{relation}', `\tt{subclass}', `\tt{restriction}', `\tt{abridgment}', `\tt{composition}', `\tt{identity}' 	
& `\tt{category}', `\tt{functor}', `\tt{natural-transformation}', `\tt{adjunction}', `\tt{monad}', `\tt{graph}', `\tt{partial-order}', `\tt{total-order}', `\tt{equivalence-relation}',   
\\ \cline{2-2}
& \raggedright `\tt{unary-function}', `\tt{binary-function}', `\tt{curry}', `\tt{hom}', `\tt{exponent}'	
& `\tt{classification}', `\tt{infomorphism}', `\tt{concept}', `\tt{concept-lattice}', `\tt{bond}'  
\\ \cline{2-3}
\ttb{meta-lower}
& \raggedright `\tt{set}', `\tt{function}', `\tt{relation}', `\tt{subset}', `\tt{restriction}', `\tt{abridgment}', `\tt{composition}', `\tt{identity}' 	
& `\tt{entity}', `\tt{relation}', `\tt{arity}', `\tt{language}', `\tt{theory}', `\tt{logic}', `\tt{language-morphism}', `\tt{theory-morphism}', `\tt{logic-morphism}'   
\\ \cline{2-2}
& \raggedright `\tt{diagram}', `\tt{colimit}', `\tt{direct-flow}', `\tt{inverse-flow}' 	
& \\ \cline{2-3}
\rmb{object level}	
& \multicolumn{2}{p{13cm}|}{`\tt{person}', `\tt{company}', `\tt{employ}', `\tt{salary}', `\tt{date}', `\tt{real-number}', \ldots}	
\\ \cline{2-3}
\end{tabular}
\end{scriptsize}
\caption{Example IFF Terms}
\label{example-terms}
\end{table}

\begin{table}
\begin{center}
\begin{scriptsize}
\fbox{\begin{tabular}{llll} 
`\tt{abridgment}'&`\tt{antisymmetric}'&`\tt{bijection}'&`\tt{binary coproduct}'\\
`\tt{binary intersection}'&`\tt{binary product}'&`\tt{binary union}'&`\tt{bond}'\\
`\tt{bonding pair}'&`\tt{categorical equivalence}'&`\tt{category}'&`\tt{class}'\\
`\tt{classification}'&`\tt{cocone}'&`\tt{coequalizer}'&`\tt{colimit}'\\
`\tt{colimit injection}'&`\tt{collection}'&`\tt{composition}'&`\tt{concept lattice}'\\
`\tt{cone}'&`\tt{constant function}'&`\tt{coproduct}'&`\tt{currying}'\\
`\tt{diagram}'&`\tt{disjoint}'&`\tt{domain}'&`\tt{empty}'\\
`\tt{epimorphism}'&`\tt{equalizer}'&`\tt{equivalence relation}'&`\tt{evaluation}'\\
`\tt{exponent}'&`\tt{function}'&`\tt{functional relation}'&`\tt{functor}'\\
`\tt{hom}'&`\tt{hypergraph}'&`\tt{identity}'&`\tt{inclusion}'\\
`\tt{infomorphism}'&`\tt{initial}'&`\tt{injection}'&`\tt{institution}'\\
`\tt{institution morphism}'&`\tt{inverse currying}'&`\tt{isomorphic}'&`\tt{isomorphism}'\\
`\tt{language}'&`\tt{limit}'&`\tt{limit projection}'&`\tt{mediator}'\\
`\tt{monad}'&`\tt{monomorphism}'&`\tt{mor2rel}'&`\tt{morphism}'\\
`\tt{natural transformation}'&`\tt{nothing}'&`\tt{null}'&`\tt{object}'\\
`\tt{one}'&`\tt{opspan}'&`\tt{order}'&`\tt{packing}'\\
`\tt{pair}'&`\tt{parallel pair}'&`\tt{partial function}'&`\tt{partial order}'\\
`\tt{pfn2ftn}'&`\tt{pfn2rel}'&`\tt{product}'&`\tt{product with}'\\
`\tt{pullback}'&`\tt{pushout}'&`\tt{reflexive}'&`\tt{relation}'\\
`\tt{restriction}'&`\tt{set}'&`\tt{signature}'&`\tt{source}'\\
`\tt{span}'&`\tt{spangraph}'&`\tt{subclass}'&`\tt{subcollection}'\\
`\tt{subordination}'&`\tt{subset}'&`\tt{surjection}'&`\tt{symmetric}'\\
`\tt{target}'&`\tt{terminal}'&`\tt{ternary coproduct}'&`\tt{ternary product}'\\
`\tt{thing}'&`\tt{three}'&`\tt{total order}'&`\tt{total relation}'\\
`\tt{transitive}'&`\tt{triple}'&`\tt{two}'&`\tt{unique function}'\\
`\tt{unique morphism}'&`\tt{unit}'&`\tt{unpacking}'&`\tt{vocabulary}'\\
`\tt{zero}'&&&\\
\end{tabular}}
\end{scriptsize}
\end{center}
\caption{Basic IFF Terms}
\label{basic-terms}
\end{table}

There are thousands of terms in the IFF.
Terms are divided into two classes, which we can call ``usable terms'' and ``supporting terms''.
An IFF term, which is defined in a particular namespace on a particular metalevel, 
is a \emph{usable} term when it is used by at least one other term in another namespace on that metalevel or on a level below that one.
An IFF term, which is defined in a particular namespace on a particular metalevel, 
is a \emph{supporting} term when it is used by another term in that same namespace.
Because of conceptual warrant,
all IFF terms should be usable or supporting, and perhaps both.
Hence, all IFF terms are necessary, but most IFF terms are ``conceptually derived''.
This means that they are a conceptual composite of two or more basic IFF terms.
An IFF term is a \emph{basic} IFF term when it is not the conceptual composite of two or more other IFF terms.
Table~\ref{basic-terms} lists some basic IFF terms. 

There are four IFF metalevels: lower, upper, top and ur.
Each metalevel services the levels below: 
the ur metalevel services the top metalevel, 
the ur and top metalevels service the upper metalevel, 
the ur, top and upper metalevels service the lower metalevel, and 
the ur, top, upper and lower metalevels service the object-level.
There is one metalanguage based at each metalevel.
Any metalanguage can be used by the meta-ontologies and ontologies at all lower levels.
These metalanguages can be defined in terms of their top-down nesting.
Each metalevel $k$ has an associated metalanguage \tts{meta}$(k)$,
which is the union of ``old'' terminology, ``specialization'' terminology and ``new'' terminology.
The old terminology is the terminology of the metalanguage \tts{meta}$(k{+}1)$ associated with the metalevel $k{+}1$ immediately above,
the specialization terminology is the terminology of the specialization of the $k{+}1$ axiomatization to the $k^{\mathrm{th}}$ metalevel,
and the new terminology is from the non-specialization axiomatization in the various meta-ontologies at metalevel $k$.

Proceeding in a top-down fashion,
the metalanguage hierarchy starts with a logical shell called the \tts{metashell}.
The \tts{metashell} enables a lisp-like first-order expression using connectives and restricted quantification.
Table~\ref{metashell-code} contains an example expression from the collection namespace of the \tts{IFF-TCO} meta-ontology that uses much of the \tts{metashell}: it uses both universal and existential restricted quantification, plus the conjunction and equivalence connectives.
In order to define the metalanguage for a particular metalevel, 
the \tts{metashell} language is typically used in two ways.
First, the typing of the terminology for a metalanguage uses the metalanguage at the next higher metalevel, 
where the \tts{metashell} expression is restricted to membership for sets, application for functions, and holds for binary relations [this is strictly category-theoretic].
Second, the meaning of the terminology for a metalanguage uses the metalanguage at the next higher metalevel 
[this includes unrestricted \tts{metashell} (quantifiers, connectives, etc.)].

\begin{table}
\begin{center}
\begin{scriptsize}
\begin{tabular}{|p{3.85in}|} \hline
\\ 
\ttbf{vlrg.set}: ``\emph{Two (very large) sets (aka collections) are isomorphic when they are connected by a (very large) bijection.}''
\begin{verbatim}
(forall (?c0 (collection ?c0) ?c1 (collection ?c1))
    (and (iff (isomorphic ?c0 ?c1)
              (exists (?f (vlrg.ftn:bijection ?f)) 
                  (and (= (source ?f) ?c0) (= (target ?f) ?c1))))
\end{verbatim} 
\\ \hline
\end{tabular}
\end{scriptsize}
\end{center}
\caption{{\ttfamily metashell} code}
\label{metashell-code}
\end{table}

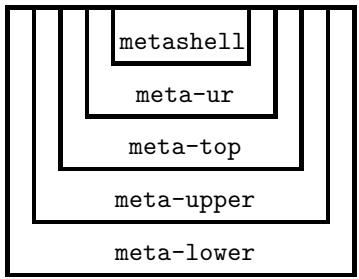
\begin{figure}
\begin{ttfamily}
\begin{bfseries}
\begin{center}
\setlength{\unitlength}{1pt}
\begin{picture}(130,110)(0,10)
\linethickness{1.3pt}
\put(40,80){\framebox(50,20){}}
\put(30,60){\framebox(70,40){}}
\put(20,40){\framebox(90,60){}}
\put(10,20){\framebox(110,80){}}
\put(0,0){\framebox(130,100){}}
\put(37,85){\makebox[2cm]{\footnotesize{metashell}}}
\put(38,65){\makebox[2cm]{\footnotesize{meta-ur}}}
\put(38,45){\makebox[2cm]{\footnotesize{meta-top}}}
\put(38,25){\makebox[2cm]{\footnotesize{meta-upper}}}
\put(38,5){\makebox[2cm]{\footnotesize{meta-lower}}}
\end{picture}
\end{center}
\end{bfseries}
\end{ttfamily}
\caption{The IFF Metalanguage Hierarchy}
\label{metalanguage-hierarchy}
\end{figure}

In addition to the \tts{metashell},
there results a hierarchy of metalanguages (Figure~\ref{metalanguage-hierarchy}) coordinated with the the metastack.
Each metalanguage includes the ones above, with the \tts{metashell} first order logical expression as innermost:
the \tts{metashell} is contained within the \tts{meta}$(4)$ = \tts{meta-ur} core metalanguage,
which is contained within the \tts{meta}$(3)$ = \tts{meta-top} metalanguage,
which is contained within the \tts{meta}$(2)$ = \tts{meta-upper} metalanguage,
which is contained within the \tts{meta}$(1)$ = \tts{meta-lower} metalanguage.
Any object-level language uses the \tts{meta-lower} metalanguage in its axiomatization.
The \tts{meta-ur} metalanguage is special --- it axiomatizes the entire metastack.
Visualizing the metalanguage nesting in terms of the metastack,
the \tts{meta-ur} metalanguage acts as a central spindle oriented vertically (see Figure~\ref{architecture-icon}), 
around which the metalanguages for the three other metalevels are ringed.

\section{The IFF Application}\label{sec-application}

Similar to software development, 
there are two approaches to the standardization of conceptual knowledge (ontology development): 
the monolithic approach and the modular approach.
But standardization strongly affects maintenance,
and in the maintenance of ontologies \cite{kent:iswc03}
the modular approach to conceptual knowledge representation has advantages over the monolithic.
In addition,
important distributed environments, such as the Semantic World-Wide Web and organizational intranets, 
represent their information in a modular fashion with multiple ontologies and schemas.
For these reasons,
the SUO advocates the modular approach in their lattice of theories (LOT) project.
The goal of the LOT project is to create a framework ``which can support an open-ended number of theories organized in a lattice together with systematic metalevel techniques for moving from one to another, for testing their adequacy for any given problem, and for mixing, matching, combining, and transforming them to whatever form is appropriate for whatever problem anyone is trying to solve'' (John F. Sowa, SUO archives, see also \cite{sowa:00}).
A challenge to the modular approach is the need for a semantic integration of ontologies.

The main application of the IFF is institutional,
involving ontology development and semantic integration.
To be logic independent,
the IFF represents and manipulates ontological structures within the metatheory of institutions
\cite{goguen:burstall:92},
which allows the formalization, representation, implementation and translation of logics.
The metatheory of institutions is being axiomatized in the upper metalevels of the IFF.
Its institutional approach to logical semantics provides a principled framework 
for the modular design of object-level ontologies in general, and 
for the ``lattice of theories'' approach to ontological organization in particular.
Within institutions, 
the lattice of theories is the fibring or indexing of the category of theories over the category of languages (aka signatures).
Institutions formally express semantic integration as an ontological fusion process \cite{kent:dagstuhl}:
align ontologies within a diagram of theories and fuse aligned ontologies via the colimit of this diagram.
The IFF has work-in-progress axiomatizations for the institutions and connecting institution morphisms
of information flow ({\bfseries IF}), equational logic ({\bfseries EQL}), order-sorted first order logic ({\bfseries FOL}) and the common logic standard ({\bfseries CL}).
Although semantic integration can be represented within an arbitrary institution \cite{kent:dagstuhl},
semantic integration in FOL-related institutions is of particular interest to the SUO project.
The architecture of FOL involves a fundamental exomorphic-endomorphic distinction\footnote{Exomorphic refers to the architecture outside the notion of an FOL language,
whereas endomorphic refers to the architecture within an FOL language.
The exomorphic aspect of FOL
is concerned with the institutional approach to FOL;
this is an indexed category architecture based on substitution along language morphisms.
The endomorphic aspect of FOL,
which is formalized in various treatments of categorical logic (see \cite{pitts:00} and \cite{seely:83}),
is concerned with the hyperdoctrine approach to FOL;
this is an indexed category architecture based on substitution along tuples in the Lawvere construction.}.
The exomorphic aspect of FOL is needed for a principled formulation of the ``lattice of theories'' concept.

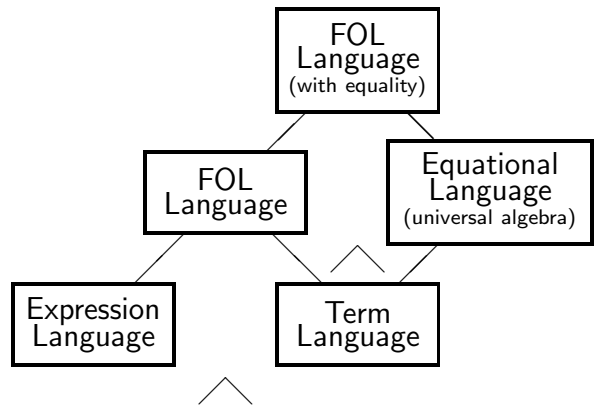
\begin{figure}
\begin{center}
\setlength{\unitlength}{1pt}
\begin{picture}(200,150)(-20,-20)
\linethickness{1.0pt}
\put(66,66){\line(1,1){14}}
\put(16,16){\line(1,1){18}}
\put(86,16){\line(-1,1){18}}
\put(116,16){\line(1,1){14}}
\put(129,70){\line(-1,1){10.5}}
\put(100,30){\line(-1,-1){10}}
\put(100,30){\line(1,-1){10}}
\put(50,-20){\line(-1,-1){10}}
\put(50,-20){\line(1,-1){10}}
\put(70,81){\framebox(60,38){}}
\put(70,91){\makebox(60,38){\sffamily{FOL}}}
\put(70,81){\makebox(60,38){\sffamily{Language}}}
\put(70,71){\makebox(60,38){{\sffamily\scriptsize{(with equality)}}}}
\put(20,35){\framebox(60,30){}}
\put(20,40){\makebox(60,30){\sffamily{FOL}}}
\put(20,30){\makebox(60,30){\sffamily{Language}}}
\put(112,31){\framebox(76,38){}}
\put(112,41){\makebox(76,38){\sffamily{Equational}}}
\put(112,31){\makebox(76,38){\sffamily{Language}}}
\put(112,21){\makebox(76,38){{\sffamily\scriptsize{(universal algebra)}}}}
\put(-30,-15){\framebox(60,30){}}
\put(-30,-10){\makebox(60,30){\sffamily{Expression}}}
\put(-30,-20){\makebox(60,30){\sffamily{Language}}}
\put(70,-15){\framebox(60,30){}}
\put(70,-10){\makebox(60,30){\sffamily{Term}}}
\put(70,-20){\makebox(60,30){\sffamily{Language}}}
\end{picture}
\end{center}
\caption{The IFF-FOL Module Hierarchy}
\label{fol-hierarchy}
\end{figure}

The IFF First Order Logic (meta) Ontology (\tts{IFF-FOL}) gives a traditional axiomatization for FOL.
The \tts{IFF-FOL} is embedded within a very modular architecture as illustrated in Figure~\ref{fol-hierarchy}.
This describes several institutions and institution morphisms,
including {\bfseries EQN} the institution for equational logic and {\bfseries FOL} the institution for first order logic (with equality).
The central bifurcation in Figure~\ref{fol-hierarchy} is between terms and expressions.
FOL languages are the pullback of expression languages and term languages over (bijections of) variables. 
FOL languages with equality are the pullback of FOL languages and universal algebra (equational languages) over term languages.
Expression languages consist of relation (type) symbols and variables.
To get FOL languages with equality, pullback relations along functions and equations.
Term languages consist of function (type) symbols and variables. 
The Lawvere construction is defined here. 
Equations can be added giving equational languages (equational presentations) as an extension of term languages. 
They define a quotient of their Lawvere category.
Expression languages consist of relation (type) symbols and variables.
Peircian existential graphs can be included here.
FOL languages consist of function (type) symbols, relation (type) symbols and variables.
From the modular perspective of Figure~\ref{fol-hierarchy},
an FOL language is a term language and an expression language that share a common set of variables. 
The important submodules in the FOL axiomatization are the following:
a term/tuple fixpoint axiomatization, an expression/arity fixpoint axiomatization, the Lawvere construction axiomatization, an axiomatization for term-tuple coproducts, the term monad axiomatization and the expression monad axiomatization. 
Each edge in the diagram of (Figure~\ref{fol-hierarchy}) is associated with institution morphisms in both directions,
projection downward and inclusion upward.

\section{Summary and Future Work}\label{sec-future-work}

In this paper, 
we have presented the Information Flow Framework (IFF) as a descriptive category metatheory
by discussing its design guidelines, development phases, architecture and institutional application.
Meta-ontologies (i.e., metatheories),
as well as dictionaries,
are classified by the prescriptive-descriptive distinction.
The development of meta-ontologies has strong analogies with the development of dictionaries.
The IFF,
which was the first of several projects accepted for ontology development under the auspices of the Standard Upper Ontology (SUO) working group,
forms its structural aspect.
The design guidelines make apparent that the IFF is being developed by following the intuitions of the working category-theorist.
The IFF architecture,
which consists of metalevels partitioned into namespaces, 
is structured by its namespace mechanism.
The IFF metastack, 
which forms the heart of the IFF architecture, 
represents the small, the large, the very large and the generic.
This is formalized in terms of four metalevels of sets, linked by functions and binary relations. 
The specialization between metalevels is effected by set subset, function restriction and relation abridgment.
Three principles discovered during the development phases of the IFF are
conceptual warrant, categorical design and institutional logic.
The linguistic aspect of the IFF is represented by four nested metalanguages,
with each metalanguage linked to a metalevel.
The metashell provides standard logical notation for the metalanguages with restricted quantification for typing.
The main application of the IFF uses the metatheory of institutions to abstractly express the modular modeling technique of the lattice of theories framework used in the development and semantic integration of object-level ontologies.

We anticipate that fibrations and indexed categories will be central to the fourth phase of IFF development.
The large categories associated with many lower level meta-ontologies are fibered.
Furthermore,
indexed categories, fibrations and the Grothendieck construction are very important to the metatheory of institutions,
whose axiomatization is currently under development in the IFF.
We also want to axiomatize sketches,
and incorporate these with the axiomatization for institutions.
Furthermore, it would be interesting to investigate
the connections between higher order logic and the first order expression at the various IFF metalevels.
Finally, we make two observations.
The IFF regards category theory as a meta-ontology,
which is exactly the spirit of the paper \cite{dampney:johnson:01}.
However, in a sense the IFF has outgrown its origins and institutional applications. 
It was originally based upon information flow and formal concept analysis with a slender category theory core. 
But now it has an increasingly strong categorically oriented super-structure.
We propose to split the old IFF into
(1) the descriptive category metatheory proper, calling this {\bfseries meta}, and
(2) a new smaller IFF that covers only the institutional application to ontology development and semantic integration.
In addition,
we propose the initiation of a working group for development of a category theory standard,
as currently manifested in {\bfseries meta}.

\tableofcontents

\bibliographystyle{plain}
\bibliography{kent}

\begin{thebibliography}{10}

\bibitem{barwise:seligman:97}
Jon Barwise and Jerry Seligman.
\newblock {\em Information Flow: The Logic of Distributed Systems}, volume~44
  of {\em Cambridge Tracts in Theoretical Computer Science}.
\newblock Cambridge University Press, 1997.

\bibitem{dampney:johnson:01}
C.N.G Dampney and Michael Johnson.
\newblock On category theory as a (meta) ontology for information systems
  research.
\newblock In Chris Welty and Barry Smith, editors, {\em Formal Ontology in
  Information Systems}, pages 59--69. ACM Press, 2001.
\newblock Proceedings of the International Conference on Formal Ontology in
  Information Systems (FOIS'01), Ogunquit, Maine.

\bibitem{ganter:wille:99}
Bernhard Ganter and Rudolf Wille.
\newblock {\em Formal Concept Analysis: Mathematical Foundations}.
\newblock Springer, 1999.
\newblock Title of the original German edition: \emph{Formale Begriffsanalyse
  --- Mathematische Grundlagen} (1996).

\bibitem{goguen:burstall:92}
Joseph Goguen and Rod Burstall.
\newblock Institutions: Abstract model theory for specification and
  programming.
\newblock {\em Journal of the Association for Computing Machinery},
  39(1):95--146, 1992.
\newblock Preprint, Report CSLI-85-30, Center for the Study of Language and
  Information, Stanford University, 1985.

\bibitem{johnson:cham}
Samuel Johnson.
\newblock {\em A Dictionary of the English Language}.
\newblock W. Strahan, 1755.

\bibitem{iff:homepage}
Robert~E. Kent.
\newblock The {SUO} {I}nformation {F}low {F}ramework ({SUO IFF}).
\newblock Technical report, {I}nstitute of {E}lectrical and {E}lectronics
  {E}ngineers, 2001.
\newblock Published as the webpage http://suo.ieee.org/IFF/].

\bibitem{kent:02}
Robert~E. Kent.
\newblock Distributed conceptual structures.
\newblock In Harre de~Swart, editor, {\em Sixth International Workshop on
  Relational Methods in Computer Science}, volume 2561 of {\em Lecture Notes in
  Computer Science}, pages 104--123. Springer, 2002.

\bibitem{kent:iswc03}
Robert~E. Kent.
\newblock Semantic integration in the {IFF}.
\newblock In AnHai Doan, Alon Halevy, and Natasha Noy, editors, {\em Semantic
  Integration 2003}, volume~82 of {\em CEUR Workshop Proceedings}. Sun SITE
  Central Europe (CEUR), 2003.
\newblock Proceedings of the Semantic Integration Workshop at ISWC-03, Sanibel
  Island, Florida, USA, October 20, 2003.

\bibitem{kent:dagstuhl}
Robert~E. Kent.
\newblock Semantic integration in the {I}nformation {F}low {F}ramework.
\newblock In Y.~Kalfoglou, M.~Schorlemmer, A.~Sheth, S.~Staab, and M.~Uschold,
  editors, {\em Semantic Interoperability and Integration}, number 04391 in
  Dagstuhl Seminar Proceedings. Dagstuhl Research Online Publication Server,
  2005.

\bibitem{lawvere:rosebrugh:03}
F.~William Lawvere and Robert Rosebrugh.
\newblock {\em Sets for Mathematics}.
\newblock Cambridge University Press, 2003.

\bibitem{lehmann:wille}
Fritz Lehmann and Rudolf Wille.
\newblock A triadic approach to formal concept analysis.
\newblock {\em Lecture Notes in Artificial Intelligence}, 954:32--43, 1995.

\bibitem{maclane:71}
Saunders {Mac Lane}.
\newblock {\em Categories for the Working Mathematician}.
\newblock Springer-Verlag, 1971.

\bibitem{m-w:col11}
Merriam-Webster, editor.
\newblock {\em Merriam-Webster's Collegiate Dictionary}.
\newblock Merriam-Webster, 11 edition, 2003.

\bibitem{goguen:logic}
Till Mossakowski, Joseph Goguen, R\u{a}van Diaconescu, and Andrzej Tarlecki.
\newblock What is a logic?
\newblock In {\em Proceedings of the First World Conference on Universal
  Logic}, 2005.

\bibitem{pitts:00}
Andrew~M. Pitts.
\newblock Categorical logic.
\newblock In S.~Abramsky, D.M. Gabbay, and T.S.E. Maibaum, editors, {\em
  Handbook of Logic in Computer Science}, volume~5 of {\em Algebraic and
  Logical Structures}. Oxford University Press, 2000.

\bibitem{suo:homepage}
James Schoening.
\newblock {IEEE} {P}1600.1 {S}tandard {U}pper {O}ntology {W}orking {G}roup
  ({SUO WG}).
\newblock Technical report, {I}nstitute of {E}lectrical and {E}lectronics
  {E}ngineers, December 2000.
\newblock Published as the webpage [http://suo.ieee.org/].

\bibitem{seely:83}
Robert~A.G. Seely.
\newblock Hyperdoctrines, natural deduction, and the beck condition.
\newblock In {\em Zeitschrift f\"{u}r Mathematische Logik und Grundlagen der
  Mathematik}, volume~29, pages 505--542, 1983.

\bibitem{sowa:00}
John~F. Sowa.
\newblock {\em Knowledge Representation: Logical, Philosophical, and
  Computational Foundations}.
\newblock Brookes/Coles, 2000.

\bibitem{winchester:03}
Simon Winchester.
\newblock {\em The Meaning of Everything: The Story of the Oxford English
  Dictionary}.
\newblock Oxford University Press, 2003.

\end{thebibliography}

\end{document}